\documentclass[12pt,a4paper]{article}
\usepackage{amsmath,amssymb} 
\usepackage{accents}
\usepackage{color}
\usepackage[bookmarks=true,hyperfigures=true]{hyperref}
\usepackage{graphicx}
\usepackage[nosort]{cite}
\usepackage[bulletsep]{collref}
\usepackage{tensor}
\usepackage{shuffle}
\usepackage{lscape}
\usepackage{empheq}
\usepackage{color}

\usepackage[a4paper,text={173mm,216mm},centering]{geometry}

\allowdisplaybreaks[3]

\usepackage[font=small,labelfont=bf,width=0.85\textwidth]{caption}

\let\oldbfseries=\bfseries
\let\oldmdseries=\mdseries
\let\oldnormalfont=\normalfont
\renewcommand{\bfseries}{\oldbfseries\boldmath}
\renewcommand{\mdseries}{\oldmdseries\unboldmath}
\renewcommand{\normalfont}{\oldnormalfont\unboldmath}

\makeatletter
\newlength{\apb@width}
\newcommand{\autoparbox}[2][c]{\settowidth{\apb@width}{#2}\parbox[#1]{\apb@width}{#2}}

\makeatother

\newcommand{\remark}[2][.]{{\color{red}\renewcommand{\bfdefault}{b}\rmfamily\if.#1\else\textbf{#1:} \fi#2}}

\newcommand{\be}{\begin{equation}}
\newcommand{\ee}{\end{equation}}
\newcommand{\beq}{\begin{equation}}
\newcommand{\eeq}{\end{equation}}

\newcommand{\bma}{\begin{pmatrix}}
\newcommand{\ema}{\end{pmatrix}}
\newcommand{\ba}{\begin{eqnarray}}
\newcommand{\ea}{\end{eqnarray}}
\newcommand{\A}{\mathcal{A}}   

\ifx\genfrac\sdflkaj\else\fi

\newcommand{\cA}{\mathcal{A}}

\newcommand{\cC}{\mathcal{C}}

\newcommand{\ep}{\epsilon}

\def\l<{\langle}\def\r>{\rangle}

\makeatletter
\newcommand{\namedref}[2]{\hyperref[#2]{#1~\ref*{#2}}}
\newcommand{\secref}{\@ifstar{\namedref{Section}}{\namedref{sec.}}}
\newcommand{\subsecref}{\@ifstar{\namedref{Subsection}}{\namedref{subsec.}}}
\newcommand{\appref}{\@ifstar{\namedref{Appendix}}{\namedref{app.}}}
\newcommand{\tabref}{\@ifstar{\namedref{Table}}{\namedref{tab.}}}
\newcommand{\figref}{\@ifstar{\namedref{Figure}}{\namedref{fig.}}}
\makeatother

\newif\ifmrnote 
\mrnotetrue
 
\usepackage[active]{srcltx}

\def\[{\begin{equation}}
\def\]{\end{equation}}
\def\<{\begin{eqnarray}}
\def\>{\end{eqnarray}}

\newcommand{\dd}{\mathrm{d}}
\newcommand{\te}{\textrm}
\newcommand{\ap}{\alpha'}

\def\bea{\begin{align}}
\def\eea{\end{align}}
\def\be{\begin{equation}}
\def\ee{\end{equation}}

\newcommand{\sv}{{\rm sv} \,}
\begin{document}
\thispagestyle{empty}
% \ifarxiv\vspace*{-20mm}\fi

\vspace{15mm}

\begin{center}

{\LARGE\bfseries Amplitude relations in heterotic string theory\par}
 
 \medskip 
 
{\LARGE\bfseries and Einstein-Yang-Mills \par}

\vspace{15mm}

\begingroup\scshape\large 
Oliver Schlotterer
\endgroup
\vspace{5mm}

\bigskip

\textit{
  Max-Planck-Institut f\"ur Gravitationsphysik, Albert-Einstein-Institut \\
  Am M\"uhlenberg 1, D-14476 Potsdam, Germany
  }  \\[0.1cm]

\bigskip
\bigskip

\texttt{olivers@aei.mpg.de} 

\vspace{14mm}

% \vspace{\fill}

\textbf{Abstract}\vspace{8mm}\par
\begin{minipage}{14.7cm}

We present all-multiplicity evidence that the tree-level S-matrix of gluons and gravitons in heterotic string theory can be reduced to color-ordered single-trace amplitudes of the gauge multiplet. Explicit amplitude relations are derived for up to three gravitons, up to two color traces and an arbitrary number of gluons in each case. The results are valid to all orders in the inverse string tension $\alpha'$ and generalize to the ten-dimensional superamplitudes which preserve 16 supercharges. Their field-theory limit results in an alternative proof of the recently discovered relations between Einstein-Yang-Mills amplitudes and those of pure Yang-Mills theory. Similarities and differences between the integrands of the Cachazo-He-Yuan formulae and the heterotic string are investigated.

 \end{minipage}\par

\end{center}
\newpage

%%%%%%%%%%%%%%%%%%%%%%%%%%%%%
% \pagenumbering{arabic} \setcounter{page}{1}
% \renewcommand{\thefootnote}{\arabic{footnote}} \setcounter{footnote}{0}

% \setcounter{tocdepth}{2} \hrule height 0.75pt
% \tableofcontents
% \vspace{0.8cm} \hrule height 0.75pt \vspace{1cm}

\setcounter{tocdepth}{2}

%%%%%%%%%%%%%%%%%%%%%%%%%%%%%%%%%%%%%%%%%%%%%%%%%%%%%%%%%%%%%%%%%%%%%%%%%%%%%%%%
%%%%%%%%%%%%%%%%%%%%%%%%%%%%%%%%%%%%%%%%%%%%%%%%%%%%%%%%%%%%%%%%%%%%%%%%%%%%%%%%

\tableofcontents

\section{Introduction}

Modern results on scattering amplitudes have revealed a fascinating interplay between perturbative gauge interactions and  gravity. A steadily growing list of examples suggests that gravitational S-matrices can be assembled from squares of suitably-arranged building blocks of amplitudes in Yang-Mills (YM) gauge theory. This double-copy structure is invisible to the field equations or a Lagrangian description but emerges naturally from string theory. Given that the massless vibration modes of open and closed strings include gauge bosons and gravitons, respectively, the double-copy construction of gravity ties in with the geometric intuition that closed strings can be formed by joining the endpoints of two open strings. For tree-level amplitudes, the Kawai--Lewellen--Tye (KLT) relations \cite{Kawai:1985xq} provided a precise prescription for squaring open-string amplitudes to those of the closed string. In the limit of vanishing string length, this immediately propagates to tree amplitudes in perturbative gauge theory and gravity. 

While a higher-genus analogue of the KLT relations is currently out of reach in string theory,
a conjectural loop-level incarnation is available in field theory \cite{Bern:2008qj, Bern:2010yg, Bern:2010ue}. The backbone of this double-copy construction is the duality between color and kinematics due to Bern, Carrasco and Johansson (BCJ) which 
organizes gauge-theory amplitudes in terms of cubic graphs and treats color and kinematic degrees of freedom on equal footing. At tree level, this duality implies the BCJ relations between color-ordered gauge-theory amplitudes \cite{Bern:2008qj} which were for instance elegantly proven from string theory \cite{BjerrumBohr:2009rd, Stieberger:2009hq}. At loop level, the BCJ duality has set new standards for multiloop calculations in various supergravities, see for instance \cite{Bern:2012uf, Bern:2012cd, Bern:2013uka, Bern:2014sna}. Explicit realizations of the kinematic constituents which obey the BCJ duality are challenging to find in general, and a variety of examples at tree level and loop level has been generated via string-theory techniques \cite{Mafra:2011kj, Mafra:2014oia, Mafra:2014gja, Mafra:2015mja, He:2015wgf,  Lee:2015upy, Mafra:2015vca}.

Less is known about mixed scattering amplitudes involving both gravitons and gauge bosons, starting with the results of \cite{Selivanov:1997ts, Selivanov:1997aq, Bern:1999bx} from the late 1990's. More recently, a double-copy description of gluon-graviton amplitudes in supergravities has been established in \cite{Chiodaroli:2014xia, Chiodaroli:2015rdg}, see \cite{Chiodaroli:2016jqw} for a review. Also here, string-theory methods provide a successful line of attack: Elegant amplitude relations for mixed open- and closed-string scattering in the type-I superstring were extracted from monodromy properties of the disk worldsheet \cite{Stieberger:2009hq, Stieberger:2015vya, Stieberger:2016lng}, valid to all orders in the inverse string tension $\alpha'$. In the field-theory limit $\alpha' \rightarrow 0$ of this setup, tree amplitudes of Einstein gravity minimally coupled to YM theory (EYM) involving $n$ gluons and a single graviton are reduced to pure gauge-theory amplitudes \cite{Stieberger:2016lng},\footnote{Throughout this work, the YM coupling $g$ as well as the gravitational constant $\kappa$ will be suppressed.}
\be
A^{\text{EYM}}(1,2,\ldots,n;p)= \sum_{l=1}^{n-1} (\epsilon_{p}\cdot x_{l})\,
A^{\text{YM}}(1,2,\ldots, l,p,l+1,\ldots,n)\, .
\label{1.0}
\ee
The cyclically ordered gluon legs $1,2,\ldots,n$ refer to a single trace of nonabelian gauge-group generators $t^{a_1} t^{a_2}\ldots t^{a_n}$, and
the graviton leg labelled by $p$ is associated with a polarization vector $\ep_p$. The latter contracts a region variable which encompasses several lightlike gluon momenta $k_{i}$,
\beq
x_{j} \equiv \sum_{i=1}^{j}k_{i} \ .
\label{1.2}
\eeq
Another approach to manifest double-copy structures is the Cachazo-He-Yuan (CHY) formalism \cite{Cachazo:2013gna, Cachazo:2013hca, Cachazo:2013iea}. Similar to string theory, tree-level amplitudes in a variety of theories and spacetime dimensions are represented as integrals over the moduli space of punctured spheres. The CHY formulae for EYM amplitudes \cite{Cachazo:2014nsa, Cachazo:2014xea} were recently exploited in \cite{Nandan:2016pya, delaCruz:2016gnm} to give an alternative proof of (\ref{1.0}), along with different sorts of generalizations to multiple graviton insertions and additional relations for double-trace amplitudes in \cite{Nandan:2016pya}.

In absence of gluon-graviton couplings, the CHY integrands for gluon and graviton scattering are essentially identical to those in the heterotic and type-II string theory, respectively. The integrands for the gauge supermultiplet in the pure-spinor incarnation of the CHY formalism \cite{Berkovits:2013xba} were shown in \cite{Gomez:2013wza} to reproduce the string-theory integrands \cite{Mafra:2011kj, Mafra:2011nv, Mafra:2011nw} derived from the open pure-spinor superstring \cite{Berkovits:2000fe}.
The CHY formulae for EYM amplitudes \cite{Cachazo:2014nsa, Cachazo:2014xea} involving both gluons and gravitons, however, do not resemble any string-theory integrands in an obvious manner. This work contributes to clarifying the relation between the CHY integrands for EYM in \cite{Cachazo:2014nsa, Cachazo:2014xea} and string theory.

In this paper, we will study gluon-graviton scattering in the heterotic string, where also the gauge multiplet arises from closed strings, and the gravity sector descends from a chiral half of the bosonic string. The heterotic integrand for tree-level amplitudes of gluons and gravitons shares the overall structure of the CHY description in \cite{Cachazo:2014nsa, Cachazo:2014xea} but differs in the contributions from the bosonic string. As we will show, the field-theory relation (\ref{1.0}) of \cite{Stieberger:2016lng} with a single graviton insertion literally carries over to single-trace amplitudes of the heterotic string,
\be
\A^{\text{het}}(1,2,\ldots,n;p)= \sum_{l=1}^{n-1} (\epsilon_{p}\cdot x_{l})\,
 \A^{\rm het}(1,2,\ldots, l,p,l+1,\ldots,n)\, ,
\label{1.1}
\ee
including the full-fledged $\alpha'$-dependence on both sides. For several gravitons, however, we will see that the expressions in the heterotic theory involve additional tensor structures in their $\alpha'$-corrections as compared to the field-theory relations from the CHY formalism \cite{Nandan:2016pya}. Two-graviton amplitudes, for instance, will be reduced to purely gluonic single-trace amplitudes via\footnote{Note that the terms with $i=1$ in the second and third line of (\ref{1.3}) yield the color ordering ${\cal A}^{\rm het}(q,1,2,\ldots,n)$ with $q$ adjacent to $n$. Moreover, color-ordered amplitudes with adjacent gravitons can be found in each line of (\ref{1.3}) including the contributions
$\sum_{i=1}^{n-1}(\ep_p\cdot x_i) $
$(\ep_q\cdot x_i)[ {\cal A}^{\rm het}(1,\ldots,i,p,q,i+1,\ldots,n)+{\cal A}^{\rm het}(1,\ldots,i,q,p,i+1,\ldots,n)]$
from the $i=j$ terms in the first line.} 
\begin{align} 
\cA^{\text{het}} (1,2, \ldots,n; p,q)\
&=  
\sum_{1=i\leq j}^{n-1}   (\epsilon_{p}\cdot x_{i})\, (\epsilon_{q}\cdot x_{j})\, 
 \cA^{\rm het}(1,\ldots,i,p,i{+}1,\ldots, j, q,j{+}1,\ldots, n)  \notag \\
&\! \! \!  \! \! \!  \! \! \!  \! \! \!  \! \! \! \! \! \! \! \! \! \! \! \! \! \! \! \! \! \! \! \! \! \! \! \! \! \! \! \! \! \! \! \! \! \! \! 
- { \big[ (\epsilon_{p}  \cdot  \epsilon_{q})  - 2\alpha'  (\epsilon_{p}  \cdot  q)  (\epsilon_{q}  \cdot p)  \big] \over 2 \, \big[1-2\alpha' (p \cdot q) \big]} \sum_{l=1}^{n-1} (p \cdot k_l)  \sum_{1=i\leq j}^l 
 \cA^{\rm het}( 1,2,\ldots, i{-}1,q,i, \ldots,j{-}1,p,j,\dots ,n) 
 \notag 
 \\
&  \! \! \!  \! \! \!  \! \! \!  \! \! \!  \! \! \! \! \! \! \! \! \! \! \! \! \! \! \! \! \! \! \! \! \! \! \! \! \! \! \! \! \! \! \! \! \! \! \! 
  - (\epsilon_{q} \cdot p)  \sum_{j=1}^{n-1}
  (\epsilon_{p}  \cdot x_{j}) \sum_{i=1}^{j+1} 
\,
\cA^{\rm het}(1,2,\ldots,i{-}1,q, i, \ldots, j,p, j{+}1,\ldots,n) 
+ (  p \leftrightarrow q) \, , \label{1.3} 
\end{align}
where the contractions $(\epsilon_{p}  \cdot  q)  (\epsilon_{q}  \cdot p)$ of the graviton momenta $p,q$ in the second line are subleading in $\alpha'$ and therefore absent in the corresponding field-theory relation \cite{Nandan:2016pya}. In the field-theory limit $\alpha' \rightarrow 0$, the relations (\ref{1.1}) and (\ref{1.3}) of the heterotic string are easily seen to reproduce the EYM relations (\ref{1.0}) due to \cite{Stieberger:2016lng} and their two-graviton counterparts given in \cite{Nandan:2016pya} and (\ref{twogravEYM}). Analogous statements apply to the amplitude relation (\ref{threegr}) involving three gravitons and $n$ gluons.

The heterotic-string approach to gluon-graviton scattering is complementary to the type-I results of \cite{Stieberger:2009hq, Stieberger:2015vya, Stieberger:2016lng} and governed by worldsheets of sphere topology instead of disks. Accordingly, our methods to derive heterotic amplitude relations (\ref{1.1}) and (\ref{1.3}) are based on integrand manipulations similar to those in the CHY formalism \cite{Nandan:2016pya} and do not involve any monodromy considerations. While multitrace amplitudes in the type-I theory appear at higher orders of the genus expansion, the heterotic string also incorporates multitrace contributions at tree level. This will be exploited to reduce color-ordered double-trace amplitudes of the heterotic string involving at most one graviton to their single-trace counterparts. The main results for the double-trace sector can be found in (\ref{4.2}) and (\ref{4.41}), and their derivation closely follows the integrand manipulations of the CHY formalism \cite{Nandan:2016pya}. 

The amplitude relations of this work also connect gluon-graviton scattering in the heterotic string with pure gluon amplitudes of the type-I string, thereby hinting an extension of the heterotic-type-I duality \cite{Polchinski:1995df} along the lines of \cite{Stieberger:2014hba}: Based on recent results on the $\alpha'$-expansion of tree-level amplitudes \cite{Schlotterer:2012ny, Stieberger:2013wea}, single-trace gluon amplitudes in the heterotic string were identified in \cite{Stieberger:2014hba} to be single-valued images of their type-I counterparts. The single-valued map truncates the multiple zeta values (MZVs) in their $\alpha'$-expansion to the subclass which can be obtained from single-valued polylogarithms \cite{Schnetz:2013hqa, Brown:2013gia}. Hence, any instance of ${\cal A}^{\rm het}(\ldots)$ on the right hand side of (\ref{1.1}), (\ref{1.3}) and later such relations may be read as $\sv {\cal A}^{\rm I}(\ldots)$, where sv denotes the single-valued map \cite{Schnetz:2013hqa, Brown:2013gia} and ${\cal A}^{\rm I}$ refers to color-ordered amplitudes (\ref{2.3}) of the type-I superstring \cite{Mafra:2011nv, Mafra:2011nw}. The availability of such connections with single-valued open-string amplitudes in presence of gravitons and multiple traces was pointed out in \cite{Stieberger:2014hba}, and we provide an explicit realization for a variety of cases.

From the pure-spinor methods in the supersymmetric side of the heterotic-string calculation \cite{Mafra:2011nv, Mafra:2011nw}, it is clear that our results including (\ref{1.1}) and (\ref{1.3}) apply to the entire superamplitudes involving multiplets of ten-dimensional SYM and half-maximal supergravity. Moreover, they are valid for any dimensional reduction of the ten-dimensional theories including any helicity sector upon specialization to four dimensions. Finally, with appropriate constraints on the external states, the results hold in all string compactifications which admit a CFT description, independently on the number of supercharges they preserve.

%%%%%%%%%%%%%%%%%%%%%%
\section{Review} 
\label{section2}
%%%%%%%%%%%%%%%%%%%%%%

In this section, we review selected aspects of tree-level amplitudes in the type-I superstring and the heterotic string. As a starting point, we recall that the vertex operators of the gauge- and gravity multiplet in the heterotic string are given by\footnote{To avoid proliferation of denominators such as $\frac{1}{2}$, we pick the abusive convention to present all factors of the inverse string tension $\alpha'$ as their appear from open strings. The exact normalizations for the heterotic string can be obtained by the replacement $\alpha' \rightarrow \frac{\alpha'}{4}$.} \cite{Gross:1985rr}
\begin{align}
{\cal V}_i^{{\rm gauge}}(z,\bar z) &\equiv {\cal J}^{a_i}(z) \, V_i^{\rm SUSY}(\bar z) \, e^{k_i\cdot X(z,\bar z)} 
\label{2.1}  \\
{\cal V}_i^{{\rm gravity}}(z,\bar z) &\equiv -\, \frac{1}{2\alpha'} \, \ep_i^\mu \, \partial X_\mu(z) \, V_i^{\rm SUSY}(\bar z) \, e^{k_i\cdot X(z,\bar z)} \ ,
\label{2.2}
\end{align}
with vector indices $\mu=0,1,\ldots,D-1$ in $D$ dimensions and worldsheet coordinates $z \in \mathbb C$. The subscript $i$ refers to the color generator $t^{a_i}$, polarization $\ep_i$, and momentum $k_i$ associated with the $i^{\rm th}$ leg. The right-moving parts $V_i^{\rm SUSY}(\bar z)$ coincide with the vertex operators of the gauge multiplet in the type-I superstring\footnote{The representation of type-I vertex operators $V^{\rm SUSY}(\bar z)$ in terms of conformal fields differs between the Ramond-Neveu-Schwarz, Green-Schwarz or pure-spinor formulation of the superstring. We will only make use of their formalism-agnostic $n$-point correlation function (\ref{2.5}) in this work and do not distinguish between integrated and unintegrated vertex operators.} 
after stripping off their color degrees of freedom, see section \ref{section21} for their correlation functions. The color-dependence in the heterotic vertex operator (\ref{2.1}) of the gauge multiplet enters through the left-moving Kac-Moody currents ${\cal J}^a(z)$. The latter are replaced by polarization-dependent operators $\ep_\mu \, \partial X^\mu(z)$ in the gravity counterpart (\ref{2.2}) subject to transversality $(\ep \cdot k)=0$. Their correlation functions relevant to this work are reviewed in section \ref{section232}. 
 
%%%%%%%%%%%%%%%%%%%%%%
\subsection{Type-I superstring} 
\label{section21}
%%%%%%%%%%%%%%%%%%%%%%

The $n$-point correlation functions of the supersymmetric vertex operators $ V^{\rm SUSY}(\bar z)$ have been evaluated in \cite{Mafra:2011nv, Mafra:2011nw} by means of the pure-spinor formalism\footnote{See \cite{Barreiro:2013dpa} for a check of the bosonic components of (\ref{2.5}) and (\ref{2.6}) at multiplicity $n\leq7$ within the RNS formalism.}
 \cite{Berkovits:2000fe}. These correlators yield the following massless amplitudes of the type-I superstring,
\begin{align}
{\cal A}^{\rm I}(1,2,\ldots,n) = \sum_{\sigma \in S_{n-3}} F^{\sigma}(\alpha') \, A^{\rm{YM}}(1,\sigma(2,\ldots,n{-}2),n{-}1,n) \ ,
\label{2.3}
\end{align}
where $S_{n-3}$ denotes the set of permutations $\sigma$ of $n-3$ legs. All the polarization dependence in (\ref{2.3}) is captured by the color-ordered tree amplitudes $A^{\rm{YM}}(\ldots)$ in ten-dimensional SYM \cite{Mafra:2010jq, Mafra:2015vca}, whereas $\alpha'$ enters through the iterated integrals $F^{\sigma}(\alpha') $ whose definition will become clear from the subsequent discussion. The structure of (\ref{2.3}) can be related to the KLT formula \cite{Kawai:1985xq} by introducing holomorphic and antiholomorphic Parke-Taylor factors
\beq
{\cal C}_z(1,2,\ldots,n) \equiv \frac{1}{z_{12} z_{23} \, \ldots \, z_{n,1}} \ , \ \ \ \ \ \
{\cal C}_{\bar z}(1,2,\ldots,n) \equiv \frac{1}{\bar z_{12} \bar z_{23} \, \ldots \, \bar z_{n,1}} \ , \ \ \ \ \ \
z_{ij} \equiv z_i - z_j
 \ .
\label{2.4}
\eeq
In this notation, the supersymmetric correlator of \cite{Mafra:2011nv, Mafra:2011nw} takes the form \cite{Broedel:2013tta}
\begin{align}
{\cal K}_n(\{\bar z_i\}) &\equiv \Big \langle \, V_1^{\rm SUSY}(\bar z_1) \, V_2^{\rm SUSY}(\bar z_2) \, \ldots \, V_n^{\rm SUSY}(\bar z_n) \, \prod_{j=1}^{n} e^{k_j \cdot X(z_j,\bar z_j)}  \, \Big \rangle 
 \label{2.5} \\
&= {\cal I}_n \sum_{\sigma,\tau \in S_{n-3}} {\cal C}_{\bar z}(1,\sigma(2,\ldots,n{-}2),n,n{-}1) \, S_0[\sigma|\tau]_1 \,
A^{\rm{YM}}(1,\tau(2,\ldots,n{-}2),n{-}1,n) \ ,
\notag
\end{align}
with the following expressions for the Koba-Nielsen factor ${\cal I}_n$ and the Mandelstam invariants $s_{ij}$,
\beq
{\cal I}_n \equiv \Big \langle \, \prod_{j=1}^{n} e^{k_j \cdot X(z_j,\bar z_j)}  \, \Big \rangle = \prod_{i<j}^n |z_{ij}|^{ 2\alpha' s_{ij}} \ , \ \ \ \ \ \
s_{ij}\equiv k_i \cdot k_j \ .
\label{2.kn}
\eeq
The $(n{-}3)! \times (n{-}3)! $ matrix $S_0[\sigma|\tau]_1$ in (\ref{2.5}) has entries of order $s_{ij}^{n-3}$ and can also be found in the $n$-point supergravity tree upon replacing $ {\cal C}_{\bar z}(\ldots)$ by another copy of $A^{\rm{YM}}(\ldots)$ \cite{Kawai:1985xq, Bern:1998sv}, see \cite{BjerrumBohr:2010ta, BjerrumBohr:2010yc, BjerrumBohr:2010hn} for its description in the momentum-kernel formalism. Note that, up to the Koba-Nielsen factor (\ref{2.kn}), the bosonic components of (\ref{2.5}) yield the reduced Pfaffian of the CHY formalism \cite{Cachazo:2013iea} while the fermionic components\footnote{A variety of examples at multiplicity $n\leq 8$ are available for download on \cite{PSwebsite}.} yield its supersymmetrization along the lines of \cite{Berkovits:2013xba, Gomez:2013wza}. Note that total derivatives in $\bar z$ have been dropped in the derivation of (\ref{2.5}) \cite{Mafra:2011nv, Mafra:2011nw}, they do not contribute to the worldsheet integrals in presence of the Koba-Nielsen factor (\ref{2.kn}).

Color-ordered amplitudes of the type-I superstring are obtained as an iterated integral
\beq
{\cal A}^{\rm I}(1,2,\ldots,n) = \int_{D(12\ldots n)} \frac{ \dd z_1 \, \dd z_2 \, \ldots \, \dd z_n }{ {\rm vol}({\rm SL}(2,\mathbb R))} \;
{\cal K}_n(\{ z_i\}) \ ,
\label{2.6}
\eeq
where the order of the gauge-group generators ${\rm Tr}(t^{a_1} t^{a_2} \ldots t^{a_n})$ in the 
accompanying trace is reflected by the integration domain
\beq
D(12\ldots n) \equiv \{ (z_1,z_2,\ldots, z_n) \in \mathbb R^n \ , \   \    -\infty < z_1 < z_2 < \ldots < z_n < \infty\} \ .
\label{2.7}
\eeq
The formal division by $ {\rm vol}({\rm SL}(2,\mathbb R))$ instructs to remove the redundancy due to Moebius transformations
$z \rightarrow \frac{ az+b}{cz+d}$ with $\left( \begin{smallmatrix} a &b \\ c &d \end{smallmatrix} \right) \in  {\rm SL}(2,\mathbb R)$. This amounts to dropping the integrations w.r.t.\ any three punctures $z_i,z_j,z_k$ and to multiply by the Jacobian $z_{ij}z_{ik} z_{jk}$. Analogous statements hold for the inverse of $ {\rm vol}({\rm SL}(2,\mathbb C))$ seen in the following section. Note that the functions $F^\sigma(\alpha')$ in (\ref{2.3}) can be assembled by inserting the correlator (\ref{2.5}) into (\ref{2.6}).

%%%%%%%%%%%%%%%%%%%%%%
\subsection{Heterotic string} 
\label{section23}
%%%%%%%%%%%%%%%%%%%%%%

Color-dressed tree-level amplitudes in heterotic string theory involving $n$ gluons and $r$ gravitons are computed from the prescription \cite{Gross:1985rr}
\beq
{\cal M}^{\rm het}(1,2,\ldots,n;n{+}1,\ldots,n{+}r) = \int_{\mathbb C^{n+r}} \! \! \! \! \frac{ \dd^2 z_1 \,  \ldots \, \dd^2 z_{n+r} }{ {\rm vol}({\rm SL}(2,\mathbb C))} \; \Big \langle \,
\prod_{i=1}^{n}  {\cal V}_i^{{\rm gauge}}(z_i,\bar z_i) \! \prod_{j=n+1}^{n+r} \! {\cal V}_{j}^{{\rm gravity}}(z_j,\bar z_j)  \, \Big \rangle \ .
\label{2.14}
\eeq
In view of the supersymmetric constituents $V^{\rm SUSY}(\bar z)$ in the vertex operators (\ref{2.1}) and (\ref{2.2}), it is clear that the type-I correlator $ {\cal K}_{n+r}(\{\bar z_j\})$ given by (\ref{2.5}) can always be factored out from the integrand. In the following, we will discuss a convenient treatment of the additional ingredients ${\cal J}^{a}$ and $\ep_\mu \partial X^\mu$ of the heterotic string which enter through the second line of
\begin{align}
&{\cal M}^{\rm het}(1,2,\ldots,n;n{+}1,\ldots,n{+}r) =  \int_{\mathbb C^{n+r}} \! \! \! \! \frac{ \dd^2 z_1 \,  \ldots \, \dd^2 z_{n+r} }{ {\rm vol}({\rm SL}(2,\mathbb C))} \; {\cal K}_{n+r}(\{\bar z_j\}) \label{2.15} \\
& \ \ \times \, \langle \, {\cal J}^{a_1}(z_1)\,  {\cal J}^{a_2}(z_2) \, \ldots  {\cal J}^{a_n}(z_n)  \, \rangle \, R_n(\ep_{n+1},\ep_{n+2},\ldots,\ep_{n+r};\{k_j,z_j\}) \ .
\notag
\end{align}
Note that the definition (\ref{2.5}) of the supersymmetric correlator ${\cal K}_{n+r}(\{\bar z_j\})$ already incorporates the Koba-Nielsen factor (\ref{2.kn}), that is why we will be interested in the reduced correlation function
\beq
R_n(\ep_{n+1},\ep_{n+2},\ldots,\ep_{n+r};\{k_j,z_j\})
\equiv \frac{(-2\alpha')^{-r}}{{\cal I}_{n+r} } \;  \Big \langle \, \prod_{j=n+1}^{n+r}\ep^\mu_j  \partial X_\mu(z_j) \prod_{i=1}^{n+r} e^{k_i \cdot X(z_i,\bar z_i)}  \, \Big \rangle \ ,
\label{2.16}
\eeq
see (\ref{2.23}) to (\ref{2.25}) for examples.

%%%%%%%%%%%%%%%%%%%%%%
\subsubsection{Current correlators and color ordering} 
\label{section231}
%%%%%%%%%%%%%%%%%%%%%%

Tree-level correlation functions among Kac-Moody currents ${\cal J}^a$ boil down to products of color traces ${\rm Tr}(t^{a_1} t^{a_2} \ldots t^{a_n}) $ accompanied by Parke-Taylor factors $(z_{12} z_{23}\ldots z_{n,1})^{-1}$ in the corresponding cyclic order. Expressions for the full correlators with a systematic discussion of their multi-trace structure can be found in \cite{Frenkel:2010ys, Dolan:2007eh}. We will mostly be interested in the single- and double-trace contributions
\begin{align}
\langle \, {\cal J}^{a_1}(z_1)\,  {\cal J}^{a_2}(z_2) \, \ldots  {\cal J}^{a_n}(z_n)  \, \rangle \, \Big|_{\te{Tr}(t^{a_1}t^{a_2} \ldots t^{a_n})} &\sim {\cal C}_z(1,2,\ldots,n) \label{2.19}
\\
\langle \, {\cal J}^{a_1}(z_1)\,  {\cal J}^{a_2}(z_2) \, \ldots  {\cal J}^{a_n}(z_n)  \, \rangle \, \Big|_{\te{Tr}(t^{a_1}t^{a_2} \ldots t^{a_p}) \atop{ \te{Tr} (t^{a_{p+1}} \ldots t^{a_n})}} &\sim {\cal C}_z(1,2,\ldots,p)  \, {\cal C}_z(p{+}1,\ldots,n)  
\label{2.20}
\end{align}
with an obvious generalization to arbitrary products of traces. Upon insertion into (\ref{2.15}), this leads to the following expressions for the color-ordered single-trace amplitudes
\begin{align}
{\cal A}^{\rm het}(1,2,\ldots,n;n{+}1,\ldots,n{+}r) &=   \int_{\mathbb C^{n+r}} \! \! \! \! \frac{ \dd^2 z_1 \,  \ldots \, \dd^2 z_{n+r} }{ {\rm vol}({\rm SL}(2,\mathbb C))} \; {\cal K}_{n+r}(\{\bar z_j\}) \label{2.21} \\
& \! \! \! \! \! \! \! \! \! \! \! \! \! \! \! \! \! \!  \times \, {\cal C}_z(1,2,\ldots,n) \, R_n(\ep_{n+1},\ep_{n+2},\ldots,\ep_{n+r};\{k_j,z_j\})
\notag
\end{align}
associated with $\te{Tr}(t^{a_1}t^{a_2} \ldots t^{a_n})$. Their double-trace counterparts
\begin{align}
{\cal A}^{\rm het}_{(2)}( \{1,2,\ldots,p\, |\, p{+}1,\ldots,n\};n{+}1,\ldots,n{+}r) &= - \int_{\mathbb C^{n+r}} \! \! \! \! \frac{ \dd^2 z_1 \,  \ldots \, \dd^2 z_{n+r} }{ {\rm vol}({\rm SL}(2,\mathbb C))} \; {\cal K}_{n+r}(\{\bar z_j\}) \label{2.22} \\
& \! \! \! \! \! \! \! \! \! \! \! \! \! \! \! \! \! \! \! \! \! \! \! \! \!  \! \! \! \! \! \! \! \! \! \! \! \! \! \! \!  \! \! \! \! \! \! \! \! \!  \! \! \! \! \! \! \! \! \! \! \!\times  \, {\cal C}_z(1,2,\ldots,p) \,  {\cal C}_z(p{+}1,\ldots,n)  \, 
R_n(\ep_{n+1},\ep_{n+2},\ldots,\ep_{n+r};\{k_j,z_j\}) \ ,
\notag
\end{align}
along with $ \te{Tr}(t^{a_1}t^{a_2} \ldots t^{a_p})\te{Tr} (t^{a_{p+1}} \ldots t^{a_n})$ involve an additional minus sign relative to the single-trace sector which can be traced back to the current correlator.

%%%%%%%%%%%%%%%%%%%%%%
\subsubsection{Single-trace gauge amplitudes as the central building block} 
\label{section239}
%%%%%%%%%%%%%%%%%%%%%%

In absence of gravitons, (\ref{2.21}) yields the following single-trace amplitudes for the gauge-multiplet
\beq
{\cal A}^{\rm het}(1,2,\ldots,n) = \int_{\mathbb C^n} \frac{ \dd^2 z_1 \, \dd^2 z_2 \, \ldots \, \dd^2 z_n }{ {\rm vol}({\rm SL}(2,\mathbb C))} \; {\cal C}_{z}(1,2,\ldots,n) \,
{\cal K}_n(\{\bar z_i\}) \ ,
\label{2.13}
\eeq
which will serve as the central building block for the remainder of this work. It will be shown to capture the supersymmetric polarizations in a variety of single- and double-trace cases (\ref{2.21}) and (\ref{2.22}) involving both gluons and gravitons. For this purpose, the second line of their integrand needs to be
expanded in $(n{+}r)$-particle Parke-Taylor factors (\ref{2.4}) such as
\begin{align}
&{\cal C}_z(1,2,\ldots,n) \, 
R_n(\ep_{n+1},\ep_{n+2},\ldots,\ep_{n+r};\{k_j,z_j\})
 =  \! \! \! \sum_{\rho \in S_{n+r-1}} \! \! \! f_\rho(\{\ep_j,k_j\})\,  {\cal C}_z(1,\rho(2,\ldots,n{+}r)) 
 \label{2.17a}\end{align}
with some functions $f_\rho(\{\ep_j,k_j\})$ of the gravitons' half-polarizations $\ep_j$ and momenta $k_j$ as well as $\alpha'$. The rearrangement in (\ref{2.17a}) will be explicitly performed in section \ref{section3} for single-trace amplitudes (\ref{2.21}) with up to three gravitons, leading to amplitude relations
\beq
{\cal A}^{\rm het}(1,2,\ldots,n;n{+}1,\ldots,n{+}r) 
=  \! \! \! \sum_{\rho \in S_{n+r-1}} \! \! \!  f_\rho(\{\ep_j,k_j\}) \, {\cal A}^{\rm het}(1,\rho(2,\ldots,n{+}r)) \ .
\label{2.17x}
\eeq
The analogous manipulations for double-trace amplitudes (\ref{2.22})
\begin{align}
&{\cal C}_z(1,2, \ldots  ,p) \, {\cal C}_z(p{+}1, \ldots  ,n) \, 
R_n(\ep_{n+1},\ldots  ,\ep_{n+r};\{k_j,z_j\})
 =  \! \! \! \sum_{\rho \in S_{n+r-1}} \! \! \! g_\rho(\{\ep_j,k_j\})\, {\cal C}_z(1,\rho(2,\ldots  ,n{+}r)) 
 \label{2.17b}\end{align}
are discussed in section \ref{section5}, where the functions $g_\rho(\{\ep_j,k_j\})$ will be explicitly determined for $r=0,1$ gravitons, and the resulting amplitude relations read
\beq
{\cal A}^{\rm het}_{(2)}( \{1,\ldots,p\, |\, p{+}1,\ldots,n\};n{+}1,\ldots,n{+}r)
 = -  \! \! \! \sum_{\rho \in S_{n+r-1}} \! \! \!  g_\rho(\{\ep_j,k_j\})\, {\cal A}^{\rm het}(1,\rho(2,\ldots  ,n{+}r))  \ .
\label{2.17y}
\eeq
Note that the rearrangements in (\ref{2.17a}) and (\ref{2.17b}) usually include integration by parts, where the total derivatives in $z$ do not contribute to the integrals of (\ref{2.15}) and will be dropped.

%%%%%%%%%%%%%%%%%%%%%%
\subsubsection{Type-I versus heterotic string and BCJ relations} 
\label{section237}
%%%%%%%%%%%%%%%%%%%%%%

For the single-trace amplitudes of the gauge multiplet, an intriguing connection between the type-I expression (\ref{2.6}) and its heterotic counterpart in (\ref{2.13}) has been found in \cite{Stieberger:2014hba}. They are related by adjusting the integration measures via
\beq
\int_{D(12\ldots n)} \frac{ \dd z_1 \, \dd z_2 \, \ldots \, \dd z_n }{ {\rm vol}({\rm SL}(2,\mathbb R))} \ \rightarrow \ \int_{\mathbb C^n} \frac{ \dd^2 z_1 \, \dd^2 z_2 \, \ldots \, \dd^2 z_n }{ {\rm vol}({\rm SL}(2,\mathbb C))} \; {\cal C}_{z}(1,2,\ldots,n) \ ,
 \label{2.dom}
 \eeq
and the $\alpha'$-expansion of the heterotic-string amplitude follows from the single-valued image 
\cite{Schnetz:2013hqa, Brown:2013gia} of the accompanying MZVs\footnote{The all-multiplicity pattern of MZVs in the $\alpha'$-expansion of the open-string amplitude (\ref{2.3}) has been described in \cite{Schlotterer:2012ny}, and machine-readable expressions for the building blocks at multiplicity $n\leq 7$ can be downloaded from \cite{MZVwebsite}. After the pioneering work in \cite{Oprisa:2005wu, Stieberger:2006te}, $\alpha'$-expansions of $n$-point disk integrals can be systematically performed via polylogarithm manipulations
\cite{Broedel:2013tta} or the Drinfeld associator \cite{Broedel:2013aza} (also see \cite{Drummond:2013vz}).
}
\cite{Stieberger:2014hba}
\beq
{\cal A}^{\rm het}(1,2,\ldots,n) = \sv {\cal A}^{\rm I}(1,2,\ldots,n) \ .
\label{svpro}
\eeq
As mentioned in the introduction, the image of the single-valued map $\te{sv}$ only comprises those MZVs which arise from single-valued polylogarithms at unity, see \cite{Schnetz:2013hqa, Brown:2013gia} or \cite{Stieberger:2013wea, Stieberger:2014hba} for further details. The identity (\ref{svpro}) can be inserted into all the amplitude relations (\ref{2.17x}) and (\ref{2.17y}) to be derived in the following, establishing an explicit connection between a variety of heterotic-string amplitudes and the type-I superstring, in lines with the general argument of \cite{Stieberger:2014hba}.

As a consequence of (\ref{svpro}), heterotic single-trace amplitudes (\ref{2.13}) were concluded in \cite{Stieberger:2014hba} to obey the BCJ relations known from field-theory amplitudes \cite{Bern:2008qj},
\beq
\sum_{l=1}^{n-1} (p \cdot x_{l})\,
 \A^{\rm het}(1,2,\ldots, l,p,l+1,\ldots,n) = 0\ ,
\label{2.bcj}
\eeq 
see (\ref{1.2}) for the region momenta $x_l$. This can also be seen from integration by parts identities among the holomorphic Parke-Taylor factors in (\ref{2.13}) \cite{Broedel:2013tta} which are equivalent to the scattering equations \cite{Cachazo:2013gna, Cachazo:2013hca, Cachazo:2013iea} when applied to Parke-Taylor factors that subtend all the particles.

%%%%%%%%%%%%%%%%%%%%%%
\subsubsection{Graviton correlators} 
\label{section232}
%%%%%%%%%%%%%%%%%%%%%%

In order to prepare the desired rewritings (\ref{2.17a}) and (\ref{2.17b}) of the heterotic-string integrands, we shall display
some examples of the reduced correlation functions (\ref{2.16}) among the worldsheet bosons $X^\mu$. Such correlators are determined by the Wick contractions
\beq
\partial X^\mu(z) \, e^{k\cdot X(0)} \sim \frac{ 2\alpha' \, k^\mu}{z}  \, e^{k\cdot X(0)} \ , \ \ \ \ \ \
\partial X^\mu(z) \, \partial X^\nu(0) \sim  \frac{2 \alpha' \, \eta^{\mu \nu} }{z^2}  \ ,
\label{2.ope}
\eeq
whose repeated application yields the expression (\ref{2.kn}) for the Koba-Nielsen factor as well as the following correlators (\ref{2.16}) of interest for amplitudes with up to three gravitons:
\begin{align} 
R_n(\ep_p;\{k_j,z_j\})
& =  \sum_{j\neq p} \frac{  (\ep_p\cdot  k_j)}{z_{j,p}}   \, 
\label{2.23} \\
R_n(\ep_p,\ep_q;\{k_j,z_j\})
& =  \sum_{j\neq p} \frac{   (\ep_p\cdot  k_j)}{z_{j,p}}  \sum_{l\neq q} \frac{ (\ep_q\cdot  k_l)}{z_{l,q}}  \, + \, \frac{ (\ep_p\cdot \ep_q)}{2\alpha' \,z_{p,q}^2} \, 
\label{2.24} \\
R_n(\ep_p,\ep_q,\ep_r;\{k_j,z_j\})
& =  \sum_{j\neq p} \frac{    (\ep_p\cdot  k_j)}{z_{j,p}}  \sum_{l\neq q} \frac{  (\ep_q\cdot  k_l) }{z_{l,q}}  \sum_{i\neq r} \frac{  (\ep_r \cdot k_i) }{z_{i,r}}  \label{2.25}  \\
& \! \! \! \!  \! \! \! \!  \! \! \! \!  \! \! \! \!  \! \! \! \!  \! \! \!   \! \! \! \!  \! \! \! \!  \! \! \! \!  \! \! \! \!  \! \! \! \!   \! \! \! \!  \!+ \, \frac{1}{2\alpha'} \, \Big\{ \frac{  (\ep_p\cdot \ep_q) }{z_{p,q}^2} \, \sum_{i\neq r} \frac{   (\ep_r \cdot k_i)}{z_{i,r}}  \, + \,  \frac{   (\ep_p\cdot \ep_r) }{z_{p,r}^2} \,\sum_{l\neq q} \frac{  (\ep_q\cdot  k_l)}{z_{l,q}}   \, + \,  \frac{ (\ep_q\cdot \ep_r) }{z_{q,r}^2} \,
 \sum_{j\neq p} \frac{       (\ep_p\cdot  k_j) }{z_{j,p}}   \, \Big\} \ .
\notag
\end{align}
Note that the summation range $\sum_{j\neq p} $ is understood to also include the graviton legs different from $p$, for instance $j\in \{1,2,\ldots,n,q,r\}$ in the three-graviton correlator (\ref{2.25}). Any contraction of the form $(\ep_p\cdot \ep_q)$ is accompanied by an inverse power of $\alpha'$, and the obvious
$r$-graviton generalization will have a maximum of $\lfloor r/2\rfloor$ such factors. Note that the right hand side of (\ref{2.23}) coincides with the diagonal elements of the $C$-matrix which constitutes the off-diagonal block of the Pfaffian in the CHY formalism \cite{Cachazo:2013hca, Cachazo:2013iea}. Since these diagonal elements have played a crucial role for the derivation of EYM relations in \cite{Nandan:2016pya, delaCruz:2016gnm}, it is not surprising that (\ref{2.23}) to (\ref{2.25}) will form the backbone of very similar derivations in heterotic string theory.

%%%%%%%%%%%%%%%%%%%%%%
\section{Single-trace amplitudes with graviton insertions} 
\label{section3}
%%%%%%%%%%%%%%%%%%%%%%

In this section, we will reduce single-trace tree-level amplitudes (\ref{2.21}) of the heterotic string involving $n$ gluons and up to three gravitons to purely gluonic single-trace trees (\ref{2.13}). We derive the amplitude relations (\ref{1.1}) and (\ref{1.3}) for one and two gravitons as well as the analogous identity (\ref{threegr}) for three gravitons. The computations closely resemble the manipulations in the CHY formalism which have been performed in \cite{Nandan:2016pya} to obtain the analogous amplitude relations for the EYM field-theory limits. Among the $\alpha'$-corrections to two- and three-graviton amplitudes, we identify a few additional tensor structures at subleading orders in $\alpha'$ which are thereby absent in the field-theory relations of \cite{Nandan:2016pya}.

%%%%%%%%%%%%%%%%%%%%%%
\subsection{One graviton} 
\label{section31}
%%%%%%%%%%%%%%%%%%%%%%

Single-trace amplitudes (\ref{2.21}) involving $n$ gluons and a single graviton at leg $n{+}1\rightarrow p$ with momentum $k_p \rightarrow p$ can be simplified by rewriting the reduced correlator (\ref{2.23}) in terms of region momenta $x_j$ defined in (\ref{1.2}),
\begin{align}
R_n(\ep_p;\{k_j,z_j\}) &=
\sum_{j=1}^{n-1} (\ep_p \cdot x_j) \, \frac{ z_{j,j+1} }{z_{j,p} z_{p,j+1}} \ .
\label{3.1}
\end{align}
We have used transversality $(\ep_p \cdot p)=0$, momentum conservation $k_n = -p - \sum_{j=1}^{n-1} k_j$ and rearranged the dependence on $z_j$, see \cite{Nandan:2016pya, delaCruz:2016gnm} for analogous manipulations in the CHY formalism. The $z$-dependence on the right hand side has the suitable form to convert an $n$-particle Parke-Taylor factor (\ref{2.4}) into its $(n{+}1)$-particle counterparts,
\begin{align}
{\cal C}_z(1,2,\ldots,n) \, R_n(\ep_p;\{k_j,z_j\}) &=
\sum_{j=1}^{n-1} (\ep_p \cdot x_j) \, {\cal C}_z(1,2,\ldots,j,p,j{+}1,\ldots,n) \ .
\label{3.2}
\end{align}
According to the general strategy discussed in section \ref{section23}, we have found a rewriting of the form
(\ref{2.17a}) which makes contact with the expression (\ref{2.13}) for heterotic single-trace amplitudes among gluons. On these grounds, (\ref{3.2}) together with (\ref{2.21}) leads to the amplitude relation (\ref{1.1}) stated in the introduction. Its
field-theory limit where ${\cal A}^{\rm het}(\ldots)\rightarrow A^{\rm YM}(\ldots)$ and ${\cal A}^{\rm het}(\ldots;p)\rightarrow A^{\rm EYM}(\ldots;p)$ reproduces the EYM relation (\ref{1.0}) derived from the type-I string \cite{Stieberger:2016lng} and later from the CHY formalism \cite{Nandan:2016pya, delaCruz:2016gnm}. Note that gauge invariance of (\ref{1.1}) under $\ep_p \rightarrow p$ follows from the BCJ relations (\ref{2.bcj}) among single-trace amplitudes in the heterotic string.

%%%%%%%%%%%%%%%%%%%%%%
\subsection{Two gravitons} 
\label{section41}
%%%%%%%%%%%%%%%%%%%%%%

As a first step towards the single-trace case with two gravitons and $n$ gluons, we express the relevant reduced correlator
in a form similar to (\ref{3.1}): With the definition
 \beq
C_{pp}'   \equiv \sum_{j=1}^{n-1} (\ep_p \cdot x_j) \, \frac{ z_{j,j+1} }{z_{j,p} z_{p,j+1}}
\label{3.3}
 \eeq
adjusted to $n$ color-ordered gluons, the correlator (\ref{2.24}) takes the form
\begin{align}
R_n(\ep_p,\ep_q;\{k_j,z_j\})
&=C_{pp}' C_{qq}' +  \frac{C_{pp}'  (\ep_q \cdot p) \, z_{p,n}}{z_{p,q} z_{q,n}}  +  \frac{   C_{qq}'  (\ep_p \cdot q) \, z_{q,n}}{z_{q,p} z_{p,n}} 
- \frac{  (\ep_p \cdot q) (\ep_q \cdot p) }{z_{p,q}^2}
+ \frac{  (\ep_p \cdot \ep_q)  }{2\alpha' z_{p,q}^2}
\label{3.4}
\end{align}
after eliminating $(\ep_p\cdot k_n)$ and $(\ep_q\cdot k_n)$ via transversality and momentum conservation. We refer to the graviton momenta via $k_p \rightarrow p$ and $k_q \rightarrow q$, respectively. It remains to
cast the product of (\ref{3.4}) with ${\cal C}_z(1,2,\ldots,n)$ into the desired form (\ref{2.17a}), and we will now discuss two kinds of techniques for this purpose.

%%%%%%%%%%%%%%%%%%%%%%
\subsubsection{Algebraic rearrangements} 
\label{section411}
%%%%%%%%%%%%%%%%%%%%%%

For all the terms in (\ref{3.4}) involving at least one factor of (\ref{3.3}), one can literally repeat the manipulations in section 4.1 and section 4.2 of \cite{Nandan:2016pya}: The $z$-dependence of $C_{pp}'$ and $C_{qq}'$ can be merged with ${\cal C}_z(1,2,\ldots,n)$ to yield an $(n{+}2)$-particle Parke-Taylor factor by iterating elementary manipulations such as
\beq
\frac{1}{z_{i,j} z_{j,k}} + \frac{1}{z_{j,i} z_{i,k}} + \frac{1}{z_{i,k} z_{k,j}} = 0 \ , \ \ \ \ \ \
z_{i,j} z_{p,q} = z_{i,p} z_{j,q} - z_{i,q} z_{j,p}  \ .
\label{3.5a}
\eeq
Intermediate steps are facilitated by a choice of ${\rm SL}(2,\mathbb C)$-frame with $z_n\rightarrow \infty$, and the frame independent form of the results is given by \cite{Nandan:2016pya}
\begin{align}
{\cal C}_z(1,2,\ldots,n) C_{pp}'  C_{qq}'  &= \sum_{i,j=1\atop i\neq j}^{n-1} (\epsilon_{p}\cdot x_{i})\,(\epsilon_{q}\cdot x_{j})\,
{\cal C}_z(1,\ldots,i,p,i{+}1,\ldots,j,q,j{+}1,\ldots,n) \label{3.5} \\ &  
\  +
\sum_{i=1}^{n-1}(\epsilon_{p}\cdot x_{i})\,  (\epsilon_{q}\cdot x_{i})\Bigl [\, 
{\cal C}_z(1,\ldots,i,p,q,i{+}1,\ldots,n) + ( p \leftrightarrow q) \, \Bigr ] \notag
 \\
%%%%
{\cal C}_z(1,\ldots,n)  \,  \frac{ C_{pp}'  (\ep_q\cdot p) \, z_{p,n}}{z_{p,q} z_{q,n}} &=  - (\epsilon_{q}\cdot p)
\sum_{i=1}^{n-1} \, (\epsilon_{p}\cdot x_{i})
 \sum_{\sigma\in \{q\} \atop{\shuffle \{1, \ldots, i\}}} 
{\cal C}_z(\sigma, p, i{+}1, \ldots, n) \ .
\label{3.6}
\end{align}
The summation range for the ordered set $\sigma$ on the right hand side of (\ref{3.6}) involves the shuffle product $\shuffle$ of $\alpha \equiv \{ a_1,a_2,\ldots,a_{i}\}$ and $\beta \equiv\{ b_1,b_2,\ldots,b_{j}\}$. It is defined to collect all permutations of the union $\alpha \cup \beta$ that preserve the individual orderings among the $a_l$ and $b_l$, which is equivalent to the recursion
\be
  \alpha \shuffle\emptyset =\emptyset\shuffle \alpha = \alpha \ , \ \ \ \ \ \
\alpha \shuffle \beta \equiv \{a_1(a_2 \ldots a_{i} \shuffle \beta)\} + \{b_1(b_2 \ldots b_{j}
\shuffle \alpha )\} \ .
\label{3.7}
\ee

%%%%%%%%%%%%%%%%%%%%%%
\subsubsection{Integration by parts} 
\label{section412}
%%%%%%%%%%%%%%%%%%%%%%

The double pole $z_{p,q}^{-2}$ in the last two terms of (\ref{3.4}) requires an integration by parts to combine with ${\cal C}_z(1,2,\ldots,n)$ and to yield $(n{+}2)$-particle Parke-Taylor factors. The idea is to discard a total derivative of the Koba-Nielsen factor (\ref{2.kn}) which simplifies in an ${\rm SL}(2,\mathbb C)$-frame where $z_n \rightarrow \infty$,
\beq
\frac{1}{ 2\ap } \, \frac{\partial}{\partial z_p} \, \frac{ {\cal I}_{n+2} }{z_{p,q}}=  \frac{ {\cal I}_{n+2}}{z_{p,q}} \, \Big\{ \frac{s_{pq} - (2\alpha')^{-1} }{z_{p,q}} + \sum_{i=1}^{n-1} \frac{ s_{pi} }{z_{p,i}} \Big\} \ ,
\label{3.10}
\eeq
and integrates to zero in the open- and closed-string measures in (\ref{2.dom}). The right hand side makes contact with the scattering equations except for the offset $s_{pq} \leftrightarrow s_{pq}-\frac{1}{2\alpha'}$. By virtue of (\ref{3.10}), the last two contributions in (\ref{3.4}) along with $(\ep_p \cdot \ep_q) -  2\alpha'  (\ep_p \cdot q) (\ep_q \cdot p)$ can be rewritten as
\begin{align}
 \frac{ {\cal C}_z(1,2,\ldots,n)  }{2\alpha' z_{p,q}^2} &= 
 \frac{  {\cal C}_z(1,2,\ldots,n) }{z_{p,q} \, (1-2\alpha' s_{pq})} \, \sum_{i=1}^{n-1} \frac{ s_{pi} }{z_{p,i}} \, \frac{z_{i,n} }{z_{q,n}} \label{3.11} \\
  &=  - \frac{ 1}{1-2\alpha' s_{pq}} \, \sum_{i=1}^{n-1} s_{pi} \! \! \!  \! \! \sum_{\sigma\in\{q,p\} \atop{ \shuffle \{1, \ldots, i{-}1 \}}} \! \! \! \! \!
 \cC_z(\sigma,i,i{+}1, \ldots, n) \ .
\notag
\end{align}
The factor of $\frac{z_{i,n} }{z_{q,n}}$ at the end of the first step has been inserted to restore covariance under ${\rm SL}(2,\mathbb C)$, and it drops out as $\frac{z_{i,n} }{z_{q,n}}\rightarrow 1$ in the frame $z_n\rightarrow \infty$ of interest to (\ref{3.10}). Up to the offset $s_{pq} \leftrightarrow s_{pq}-\frac{1}{2\alpha'}$, the manipulations in (\ref{3.11}) have also been performed in section 4.3 of \cite{Nandan:2016pya}. 
In view of its repeated appearance in the three-graviton counterpart, we introduce the shorthand
\begin{align}
{\cal E}_{ij} &\equiv \frac{ (\ep_i \cdot \ep_j) - 2\alpha' (\ep_i \cdot k_j) (\ep_j \cdot k_i) }{1-2\alpha' s_{ij}}  \label{defee} \\
&=  (\ep_i \cdot \ep_j) + \big[ s_{ij}   (\ep_i \cdot \ep_j) -  (\ep_i \cdot k_j) (\ep_j \cdot k_i)\big] \sum_{m=1}^{\infty} (2\alpha')^m s_{ij}^{m-1} \  \notag
\end{align}
for the ubiquitous combination arising from (\ref{3.11}). In passing to the second line, we have expanded the denominator as a geometric series, and all the $\alpha'$-corrections are seen to involve the gauge invariant linearized field strength $ s_{ij}   (\ep_i \cdot \ep_j) -  (\ep_i \cdot k_j) (\ep_j \cdot k_i)$. By (\ref{3.11}) and (\ref{defee}), the net contribution from the double-pole part of (\ref{3.4}) is given by
\begin{align}
&{\cal C}_z(1,2,\ldots,n)  \,  \frac{ (\ep_p \cdot \ep_q) -  2\alpha'  (\ep_p \cdot q) (\ep_q \cdot p)  }{2\alpha' z_{p,q}^2} =
- \,   {\cal E}_{pq}  \, \sum_{i=1}^{n-1} s_{pi} \! \! \!   \sum_{\sigma\in\{q,p\} \atop{ \shuffle \{1, \ldots, i{-}1 \}}} \! \! \!
 \cC_z(\sigma,i,i{+}1, \ldots, n)  \ .
\label{3.11a}
\end{align}
Note that the same rearrangements could have been performed with $p$ and $q$ exchanged, and the hidden symmetry of the right hand side of (\ref{3.11a}) under $p\leftrightarrow q$ follows from further integrations by parts or from BCJ relations (\ref{2.bcj}) among the resulting ${\cal A}^{\rm het}(\ldots)$. We will explicitly symmetrize the final form for the amplitude relation in $p$ and $q$.

%%%%%%%%%%%%%%%%%%%%%%
\subsubsection{Assembling the result} 
%%%%%%%%%%%%%%%%%%%%%%

With the right hand sides of (\ref{3.5}), (\ref{3.6}) and (\ref{3.11a}), all the left-moving contributions to the two-graviton amplitude have been brought into the desired form (\ref{2.17a}). Hence, we obtain the following amplitude relation via (\ref{2.17x})
\begin{align} 
&\cA^{\text{het}} (1, \ldots,n; p,q) 
=
\sum_{1=i\leq j}^{n-1}   (\epsilon_{p}\cdot x_{i})\, (\epsilon_{q}\cdot x_{j})\, 
\cA^{\rm het}(1,\ldots,i,p,i{+}1,\ldots, j, q,j{+}1,\ldots, n)   \label{twograv} \\
&\  - \sum_{i=1}^{n-1}  \Big\{
(\epsilon_{p} \! \cdot \! x_{i})\, (\epsilon_{q} \! \cdot \! p) 
 \! \! \! \sum_{\sigma\in \{q\} \atop{\shuffle \{1, \ldots, i\}}} \! \! \! 
\cA^{\rm het}(\sigma, p, i{+}1, \ldots, n)
+ 
{ s_{pi} \, {\cal E}_{pq} \over 2}   \! \! \! \! \!
 \sum_{\sigma\in\{ q, p\} \atop{ \shuffle \{1, \ldots, i{-}1\}}} \! \! \! \! \!
 \cA^{\rm het}( \sigma, i, i{+}1, \ldots ,n) 
  \Big\} + (  p \leftrightarrow q) \, ,
\notag
\end{align}
in agreement with (\ref{1.3}) upon expansion of the sums over shuffles. With two external gluons and gravitons, for instance, (\ref{twograv}) reduces to
\begin{align}
{\cal A}^{\rm het}(1, 2; 3, 4) &=
(\ep_3\cdot k_4)(\ep_4 \cdot x_1) {\cal A}^{\rm het}(1,2,3,4)
+(\ep_4\cdot k_3)(\ep_3 \cdot x_1) {\cal A}^{\rm het}(1,2,4,3) \notag \\
&
-(\ep_3\cdot x_1)(\ep_4 \cdot x_1) {\cal A}^{\rm het}(1,3,2,4)
- s_{14} {\cal E}_{34} {\cal A}^{\rm het}(1,2,3,4) \, . \label{ex2plus2}
\end{align}

%%%%%%%%%%%%%%%%%%%%%%
\subsubsection{Tachyon suppression and transcendentality properties} 
\label{section413}
%%%%%%%%%%%%%%%%%%%%%%

Note that $\alpha'$ enters (\ref{twograv}) and (\ref{ex2plus2}) in two different places, firstly
through the geometric series (\ref{defee}) represented by ${\cal E}_{pq}$ and secondly through the expansion of $ \cA^{\rm het}(\ldots)$ on the right hand side. By (\ref{svpro}) \cite{Stieberger:2014hba}, the latter descends from the type-I results of \cite{Mafra:2011nv, Mafra:2011nw, Schlotterer:2012ny, Broedel:2013tta, Broedel:2013aza}, where the powers of $\alpha'$ tie in with the transcendental weights of the accompanying MZVs. This property of gluonic single-trace amplitudes of the heterotic string is known as {\em uniform transcendentality}.

The series representation of ${\cal E}_{pq}$ in (\ref{defee}) obviously violates uniform transcendentality, and so does any single-trace amplitude with $r\geq 2$ graviton insertions, starting with (\ref{twograv}). Only the one-graviton case (\ref{1.1}) preserves uniform transcendentality which may be attributed to the particularly simple tensor structures. This ties in with the breakdown of uniform transcendentality in the bosonic string and its implications for purely gravitational amplitudes in the heterotic string \cite{Huang:2016tag}.

One might be worried that the pole of ${\cal E}_{pq} \sim (1-2\alpha' s_{pq})^{-1}$ at $(k_p+k_q)^2=\alpha'$ signals tachyon propagation. However, it is clear from level matching that such poles cannot occur, and their cancellation rests on the compensating zeros of the  single-trace amplitudes as $(k_p+k_q)^2\rightarrow \alpha'$. In the above four-point example (\ref{ex2plus2}), the pole of ${\cal E}_{34} \sim (1-2\alpha' s_{34})^{-1}$ coincides with a zero of
\beq
 {\cal A}^{\rm het}(1,2,3,4)  = A^{\rm YM}(1,2,3,4) \, \frac{ \Gamma(1+2\alpha' s_{12} ) \Gamma(1+2\alpha' s_{13} ) \Gamma(1+2\alpha' s_{23} )  }{\Gamma(1-2\alpha' s_{12} ) \Gamma(1-2\alpha' s_{13} ) \Gamma(1-2\alpha' s_{23} ) } 
\eeq
due to the factor of $\Gamma(1-2\alpha' s_{12} )^{-1}$.

In the field-theory limit, the two $\alpha'$-dependent quantities on the right hand side of (\ref{twograv}) reduce to $ \cA^{\rm het} \rightarrow A^{\rm YM}$ and ${\cal E}_{pq} \rightarrow (\ep_p \cdot \ep_q)$, respectively, and we are left with the EYM amplitude relation
\begin{align} 
&A^{\text{EYM}} (1, 2,\ldots,n; p,q) 
=
\sum_{1=i\leq j}^{n-1}   (\epsilon_{p}\cdot x_{i})\, (\epsilon_{q}\cdot x_{j})\, 
A^{\rm YM}(1,\ldots,i,p,i{+}1,\ldots, j, q,j{+}1,\ldots, n)   \label{twogravEYM} \\
&\!  - \sum_{i=1}^{n-1}  \Big\{
(\epsilon_{p} \! \cdot \! x_{i}) (\epsilon_{q} \! \cdot \! p) 
 \! \! \! \! \sum_{\sigma\in \{q\} \atop{\shuffle \{1, \ldots, i\}}} \! \! \! \! \!
A^{\rm YM}(\sigma, p, i{+}1,\! \ldots\!, n)
+ 
{ s_{pi} \, (\epsilon_{p} \! \cdot \! \epsilon_{q}) \over 2}   \! \! \! \! \! \!
 \sum_{\sigma\in\{ q, p\} \atop{ \shuffle \{1, \ldots, i{-}1\}}} \! \! \! \! \!\! \!
A^{\rm YM}( \sigma, i, i{+}1, \!\ldots\! ,n) 
  \Big\} + (  p \leftrightarrow q) \, ,
\notag
\end{align}
known from \cite{Nandan:2016pya}.

%%%%%%%%%%%%%%%%%%%%%%
\subsection{Three gravitons} 
\label{section42}
%%%%%%%%%%%%%%%%%%%%%%

For three gravitons and $n$ gluons, the reduced correlator (\ref{2.25}) takes the form
\begin{align}
&R_n(\ep_p,\ep_q,\ep_r;\{k_j,z_j\}) = \frac{1}{2\alpha'} \, \Big\{ \, \frac{ (\ep_p \cdot \ep_q) }{z_{p,q}^2} \, \Big( C_{rr}' + \frac{ (\ep_r\cdot q) \, z_{q,n}}{z_{q,r} z_{r,n} } + \frac{ (\ep_r\cdot p) \, z_{p,n}}{z_{p,r} z_{r,n} }  \Big)+ {\rm cyc}(p,q,r) \, \Big\} 
\label{3.70}
\\
&\! + \Big( C_{rr}' \!+\! \frac{ (\ep_r\!\cdot \!q) \, z_{q,n}}{z_{q,r} z_{r,n} } + \frac{ (\ep_r\!\cdot \!p) \, z_{p,n}}{z_{p,r} z_{r,n} }  \Big)  \! \Big( C_{qq}' \!+\! \frac{ (\ep_q\!\cdot \!p) \, z_{p,n}}{z_{p,q} z_{q,n} } + \frac{ (\ep_q\!\cdot \!r) \, z_{r,n}}{z_{r,q} z_{q,n} }  \Big)  \!\Big( C_{pp}'\! +\! \frac{ (\ep_p\!\cdot \!q) \, z_{q,n}}{z_{q,p} z_{p,n} } + \frac{ (\ep_p\!\cdot \! r) \, z_{r,n}}{z_{r,p} z_{p,n} }  \Big) \ ,\notag 
\end{align}
which serves as a convenient starting point towards amplitude relations. One can identify nine tensor structures which are inequivalent under permutations of the graviton polarizations $\ep_p,\ep_q,\ep_r$ and momenta $p,q,r$:
\beq
\begin{array}{rlrlrl}
{\rm (i)} \ &C'_{pp} C_{qq}' C_{rr}' 
&{\rm (iv)} \ &C'_{pp} (\ep_q\cdot r) (\ep_r\cdot p) \phantom{gap}
&{\rm (vii)} \ &(\ep_r\cdot q) (\ep_p\cdot \ep_q)
\\
{\rm (ii)} \ &C'_{pp} C_{qq}' (\ep_r\cdot q)
&{\rm (v)} \ &C'_{rr} (\ep_p\cdot \ep_q)
&{\rm (viii)} \ &(\ep_r \cdot q)(\ep_p \cdot q)(\ep_q \cdot p) 
\\
{\rm (iii)} \ &C'_{pp} (\ep_q\cdot p) (\ep_r\cdot p) \phantom{gap}
&{\rm (vi)} \ &C'_{rr}(\ep_p \cdot q)(\ep_q \cdot p) 
&{\rm (ix)} \ &(\ep_p \cdot q)(\ep_q \cdot r)(\ep_r \cdot p) 
\end{array}
\label{tens}
\eeq
The three classes of tensor structures (vi), (viii) and (ix) were observed to drop out from the field-theory relation found in \cite{Nandan:2016pya}, and we will identify them to be subleading in $\alpha'$ similar to the terms $(\ep_p\cdot q)(\ep_q\cdot p)$ in the two-graviton case. In the following subsections, the reduction (\ref{2.17a}) to $(n{+}3)$-particle Parke-Taylor factors will be performed for all of (i) to (ix).

%%%%%%%%%%%%%%%%%%%%%%
\subsubsection{Tensor structures (i) to (iv)} 
\label{section331}
%%%%%%%%%%%%%%%%%%%%%%

When comparing the CHY integrand of \cite{Cachazo:2014nsa} with the heterotic correlator (\ref{3.70}), the tensor structures (i) to (iv) in (\ref{tens}) turn out to occur with the same functions of the punctures. Moreover, these cases do not exhibit any double poles in the $z_{i,j}$, so the subtleties of integration by parts seen in section \ref{section412} are absent, and we directly quote the results of sections 5.3 to 5.5 in \cite{Nandan:2016pya}:
\begin{align}
&{\cal C}_z(1,2,\ldots,n)R_n(\ep_p,\ep_q,\ep_r;\{k_j,z_j\}) \, \big|_{{\rm (i) \phantom{,} to \phantom{,} (iv)}} =  (\ep_p \cdot (q{+}r)) ( \ep_q \cdot r) (\ep_r \cdot x_j)  \! \! \! \sum_{\sigma \in \{p,q\} \atop{\shuffle \{ 1,2,\ldots, j\}} }  \! \! \!{\cal C}_z(\sigma,r,j{+}1,\ldots,n) \notag\\
&\ \ \ + \sum_{1=i\leq j\leq k}^{n-1}  (\ep_p \cdot x_i)  (\ep_q \cdot x_j)(\ep_r \cdot x_k)  {\cal C}_z(1,2,\ldots,i,p,i{+}1,\ldots,j,q,j{+}1,\ldots,k,r,k{+}1,\ldots,n)  \notag \\
&\ \ \ - (\ep_r\cdot p) \Big\{   \sum_{1=i\leq j}^{n-1}  (\ep_p \cdot x_i)  (\ep_q \cdot x_j)  \sum_{\sigma \in \{r\} \atop{\shuffle \{ 1,2,\ldots, i\}} } {\cal C}_z(\sigma,p,i{+}1,\ldots,j ,q,j{+}1,\ldots,n) 
  \label{3.71} \\
& \ \ \ \ \ \ \ \ \ \ \ \ \ \ \   \ \ \ \ \ +\sum_{1=j\leq i}^{n-1}  (\ep_p \cdot x_i)  (\ep_q \cdot x_j) \! \! \! \! \! \sum_{\sigma \in \{r\} \atop{\shuffle \{ 1,\ldots,j,q,j+1,\ldots, i\}} } \! \! \! \! \! {\cal C}_z(\sigma,p,i{+}1,\ldots,n) 
\Big\}  
 +{\rm perm}(p,q,r)  \ ,\notag 
\end{align}
This follow from iteration of the basic identities (\ref{3.5a}) and will not be repeated here.

%%%%%%%%%%%%%%%%%%%%%%
\subsubsection{Tensor structures (v) and (vi)} 
\label{section332}
%%%%%%%%%%%%%%%%%%%%%%

It turns out that the representatives $(\ep_p\cdot  \ep_q)C_{rr}'$ and $(\ep_p  \cdot q)(\ep_q \cdot p)C_{rr}'$ of the two tensor structures (v) and (vi) in (\ref{tens}) multiply the same worldsheet functions (including a double pole in $z_{p,q}$) and conspire to ${\cal E}_{pq}$ defined in (\ref{defee}). We can apply the integration by parts identity (\ref{3.10}) with an additional leg $j$ inserted into the Parke-Taylor factor ${\cal C}_z(1,2,\ldots,n)$:
\begin{align}
&{\cal C}_z(1,2,\ldots,n)R_n(\ep_p,\ep_q,\ep_r;\{k_j,z_j\}) \, \big|_{{\rm (v) + (vi)}} \notag \\
& \ \ \ = {\cal C}_z(1,2,\ldots,n) \Big( \frac{(\ep_p\cdot  \ep_q)}{2\alpha' \, z_{p,q}^2} - \frac{(\ep_p  \cdot q)(\ep_q \cdot p)}{z_{p,q}^2} \Big) \! \sum_{j=1}^{n-1} \frac{ (\ep_r \cdot x_j) \, z_{j,j+1}}{z_{j,r} z_{r,j+1}} + {\rm cyc}(p,q,r) \notag \\
& \ \ \ = \frac{ {\cal E}_{pq} }{2\alpha'} \, \frac{ 1 - 2\alpha' s_{pq} }{z_{p,q}^2} \,  \sum_{j=1}^{n-1} (\ep_r \cdot  x_j) \, {\cal C}_z(1,2,\ldots,j,r,j{+}1,\ldots,n) + {\rm cyc}(p,q,r)  \notag \\
&\ \ \ =  - {\cal E}_{pq} \,  \sum_{j=1}^{n-1} (\ep_r \cdot  x_j) \,  \Big\{   \sum_{i=1}^j s_{ip} \! \! \! \! \!  \! \! \sum_{\sigma \in \{q,p\} \atop{\shuffle \{ 1,2,\ldots, i{-}1\}} } \! \! \! \! \! \! \! {\cal C}_z(\sigma,i,i{+}1,\ldots,j,r,j{+}1,\ldots,n)
  \label{3.72} \\
& \ \ \ \ \ \ \ \ \  +s_{pr} \! \! \! \! \sum_{\sigma \in \{q,p\} \atop{\shuffle \{ 1,2,\ldots, j\}} } \! \! \! \! {\cal C}_z(\sigma,r,j{+}1,\ldots,n)+ \sum_{i=j+1}^{n-1} s_{ip} \! \! \! \! \!\! \! \! \! \! \sum_{\sigma \in \{q,p\} \atop{\shuffle \{ 1,\ldots,j,r,j{+}1,\ldots i-1\}} }\! \! \! \! \! \! \! \! \! \! {\cal C}_z(\sigma,i,i{+}1,\ldots,n) \Big\} + {\rm cyc}(p,q,r) \ . \notag
\end{align}
Note that the result closely resembles section 5.2 of \cite{Nandan:2016pya} up to the $\alpha'$-corrections in $(\ep_p \cdot \ep_q) \rightarrow {\cal E}_{pq}$.

%%%%%%%%%%%%%%%%%%%%%%
\subsubsection{Tensor structures (vii) and (viii)} 
\label{section333}
%%%%%%%%%%%%%%%%%%%%%%

The representatives $(\ep_p\cdot  \ep_q) (\ep_r\cdot q)$ and $(\ep_p  \cdot q)(\ep_q \cdot p) (\ep_r\cdot q)$
of the tensor structures (vii) and (viii) in (\ref{tens}) turn out to involve the same dependence on the punctures and again combine to ${\cal E}_{pq}$:
\begin{align}
&{\cal C}_z(1,2,\ldots,n)R_n(\ep_p,\ep_q,\ep_r;\{k_j,z_j\}) \, \big|_{{\rm (vii) + (viii)}} \notag \\
& \ \ \ = {\cal C}_z(1,2,\ldots,n) \Big( \frac{(\ep_p\cdot  \ep_q)}{2\alpha' \, z_{p,q}^2} - \frac{(\ep_p  \cdot q)(\ep_q \cdot p)}{z_{p,q}^2} \Big)  \, \frac{ (\ep_r\cdot q) \, z_{q,n} }{z_{q,r} z_{r,n}}  + {\rm perm}(p,q,r) \notag \\
& \ \ \ = \frac{ {\cal E}_{pq} }{2\alpha'} \,\frac{1 - 2\alpha' s_{pq}}{z_{p,q}^2}  \,  {\cal C}_z(1,2,\ldots,n) \, \frac{ (\ep_r\cdot q) \, z_{q,n} }{z_{q,r} z_{r,n}} + {\rm perm}(p,q,r)  \label{3.74} \\
& \ \ \ ={\cal E}_{pq}  (\ep_r\cdot q) \, \Big\{ - \frac{ s_{pr} \, {\cal C}_z(1,2,\ldots,n) }{z_{p,q} z_{q,r} z_{r,p} } +  \sum_{j=1}^{n-1} s_{jp}\sum_{\sigma \in \{r,q,p\} \atop{\shuffle \{ 1,2,\ldots, j-1\}} } {\cal C}_z(\sigma,j,j{+}1,\ldots,n)  \Big\} + {\rm perm}(p,q,r) \ .\notag
\end{align}
In passing to the last line, we have applied the integration by parts identity
\beq
\frac{ \partial }{\partial z_p}  \, \frac{ {\cal I}_{n+3} }{z_{p,q} z_{q,r} } = 2\alpha' \,  \frac{ {\cal I}_{n+3} }{z_{p,q} z_{q,r} } \, \Big\{ \frac{ s_{pq} - (2\alpha')^{-1} }{z_{p,q} } + \frac{ s_{pr} }{z_{p,r}} + \sum_{i=1}^{n-1} \frac{ s_{pi} }{z_{p,i}} \Big\}\ ,
\label{3.73}
\eeq
which is valid in the frame where $z_n \rightarrow \infty$ and slightly generalizes (\ref{3.10}). Note that the analogous scattering equation has been exploited in section 5.1 of \cite{Nandan:2016pya}, where the factor of $(z_{p,q} z_{q,r} z_{r,p})^{-1}$ in the last line is cancelled by other parts of the CHY integrands which do not have any heterotic-string counterpart.

%%%%%%%%%%%%%%%%%%%%%%
\subsubsection{Tensor structures (ix)} 
\label{section334}
%%%%%%%%%%%%%%%%%%%%%%

It is convenient to combine the last tensor structures (ix) in (\ref{tens}) with the terms $\sim (z_{p,q} z_{q,r} z_{r,p} )^{-1}$ in the last line of (\ref{3.74}):
\begin{align}
&{\cal C}_z(1,2,\ldots,n)\Bigg[ R_n(\ep_p,\ep_q,\ep_r;\{k_j,z_j\}) \, \big|_{{\rm (ix)}} - \Big\{ \frac{ {\cal E}_{pq}  (\ep_r\cdot q)  s_{pr} }{z_{p,q} z_{q,r} z_{r,p} }  +  {\rm perm}(p,q,r) \Big\}  \Bigg] \notag \\ 
& \ \ \ =  (1-2\alpha' s_{pqr}){\cal F}_{pqr} \,  {\cal C}_z(1,2,\ldots,n)  \, {\cal C}_z(p,q,r) \label{3.75}
\\
& \ \ \ =   2\alpha' \, {\cal F}_{pqr}   \, \sum_{j=1}^{n-1} \Big\{  s_{rj} \! \! \! \! \sum_{\sigma \in \{ p,q,r\} \atop{ \shuffle \{1,2,\ldots,j-1\} }} \! \! \! \! {\cal C}_z(\sigma,j,j{+}1,\ldots,n) - s_{qj} \! \! \! \!  \sum_{\sigma \in \{ p,r,q\} \atop{ \shuffle \{1,2,\ldots,j-1\} }}\! \! \! \!  {\cal C}_z(\sigma,j,j{+}1,\ldots,n) \Big\}
\notag
\end{align}
In passing to the second line, the tensor structures (ix) and ${\cal E}_{pq}  (\ep_r\cdot q)s_{pr}$ have been combined to yield
the totally antisymmetric and gauge invariant generalization of (\ref{defee}),
\begin{align}
{\cal F}_{pqr} &\equiv \frac{ (\ep_q \!\cdot \!p)(\ep_p\! \cdot \!r)(\ep_r \!\cdot \!q) - (\ep_p\! \cdot \!q)(\ep_q \!\cdot \!r)(\ep_r \!\cdot \!p)+ \big[ {\cal E}_{pq} ( s_{qr} (\ep_r \!\cdot\! p) - s_{pr} (\ep_r\! \cdot \!q) )  + {\rm cyc}(p,q,r)\big] }{1-2\alpha' s_{pqr}} \ ,
\label{3.76}
\end{align}
which reproduces the Lorentz trace of three linearized field strengths at the leading order in $\ap$. In passing to the third line of (\ref{3.75}), we are anticipating the integration by parts identity (\ref{4.18}), see the later section \ref{section51} for its derivation. Moreover, the following definition of multiparticle Mandelstam invariants has been used:
\beq
s_{12\ldots k} \equiv \sum_{i<j}^k s_{ij} \ .
\label{multmand}
\eeq
Note that the entire contribution in (\ref{3.75}) is subleading in $\alpha'$ and does not have any analogue in EYM and the CHY computation of \cite{Nandan:2016pya}. Moreover, permutation invariance has been obscured in the last step of (\ref{3.75}), but it can be checked via integration by parts or BCJ relations and will be reinstated by manual averaging in the final result.

%%%%%%%%%%%%%%%%%%%%%%
\subsubsection{Assembling the result} 
\label{section335}
%%%%%%%%%%%%%%%%%%%%%%

Assembling the contributions from (\ref{3.71}), (\ref{3.72}), (\ref{3.74}) as well as (\ref{3.75}) and recalling the conversion from (\ref{2.17a}) to (\ref{2.17x}), we arrive at the following amplitude relation which reduces three-graviton single-trace trees to purely gluonic ones,
\begin{align}
 &\cA^{\text{het}} (1,2, \ldots,n; p,q,r) = {\cal E}_{pq} (\ep_r \cdot q) \sum_{j=1}^{n-1} s_{jp}\sum_{\sigma \in \{r,q,p\} \atop{\shuffle \{ 1,2,\ldots, j-1\}} } {\cal A}^{\text{het}} (\sigma,j,j{+}1,\ldots,n) \notag \\
%%%%%% The b) part
&-\, \frac{  {\cal E}_{pq} }{2} \sum_{j=1}^{n-1}  (\ep_r \cdot x_j)  \Big\{
\sum_{i=1}^j s_{ip} \! \! \! \! \! \sum_{\sigma \in \{q,p\} \atop{\shuffle \{ 1,2,\ldots, i-1\}} } \! \! \! \! \! {\cal A}^{\text{het}} (\sigma,i,i{+}1,\ldots,j,r,j{+}1,\ldots,n)
\notag  \\
& \ \ \ \ \ \  \ \ \ \ \ \  + s_{pr} \sum_{\sigma \in \{q,p\} \atop{\shuffle \{ 1,2,\ldots, j\}} } {\cal A}^{\text{het}}(\sigma,r,j{+}1,\ldots,n)
+ \sum_{i=j+1}^{n-1} s_{ip} \! \! \! \! \!\! \! \! \! \! \sum_{\sigma \in \{q,p\} \atop{\shuffle \{ 1,\ldots,j,r,j{+}1,\ldots i-1\}} }\! \! \! \! \! \! \! \! \! \! {\cal A}^{\text{het}} (\sigma,i,i{+}1,\ldots,n) \Big\} \notag \\
%%%%%% The e) part
& -  ( \ep_r \cdot p) \Big\{
\sum_{1=j\leq i}^{n-1}  (\ep_p \cdot x_i)  (\ep_q \cdot x_j) \! \! \! \! \! \sum_{\sigma \in \{r\} \atop{\shuffle \{ 1,\ldots,j,q,j{+}1,\ldots, i\}} } \! \! \! \! \! {\cal A}^{\text{het}} (\sigma,p,i{+}1,\ldots,n) 
  \label{threegr} \\
& \ \ \ \ \ \  \ \ \ \ \ \ + \sum_{1=i\leq j}^{n-1}  (\ep_p \cdot x_i)  (\ep_q \cdot x_j)  \sum_{\sigma \in \{r\} \atop{\shuffle \{ 1,2,\ldots, i\}} } {\cal A}^{\text{het}} (\sigma,p,i{+}1,\ldots,j ,q,j{+}1,\ldots,n) 
\Big\} \notag \\
%%%%%% The f) part
&+\sum_{1=i\leq j\leq k}^{n-1}  (\ep_p \cdot x_i)  (\ep_q \cdot x_j)(\ep_r \cdot x_k)    {\cal A}^{\text{het}} (1,2,\ldots,i,p,i{+}1,\ldots,j,q,j{+}1,\ldots,k,r,k{+}1,\ldots,n)\notag \\
%%%%%% The cd) part
& + (\ep_p \cdot (q+r)) ( \ep_q \cdot r) \sum_{j=1}^{n-1}  (\ep_r \cdot x_j)  \sum_{\sigma \in \{p,q\} \atop{\shuffle \{ 1,2,\ldots, j\}} } {\cal A}^{\text{het}} (\sigma,r,j{+}1,\ldots,n)  \notag \\
%%%%%% The 3-cycle
&+ \frac{2 \alpha'}{3}  \, {\cal F}_{pqr}  \, 
 \sum_{j=1}^{n-1} s_{jr} \! \! \!  \sum_{ \sigma \in \{ p,q,r\} \atop { \shuffle \{ 1,2,\ldots,j-1 \} } } \! \! \!  {\cal A}^{\text{het}} (\sigma,j,j{+}1,\ldots,n)
  +{\rm perm}(p,q,r)
 \ .\notag
\end{align}
The simplest example involves two gluons and three gravitons,
\begin{align}
{\cal A}^{\rm het}(1, 2; 3, 4,5) &= {\cal E}_{35}  (\ep_4 \cdot k_5)  s_{13} {\cal A}^{\rm het}( 1, 2, 4, 5, 3)  -  (\ep_4 \!\cdot \!(k_3+k_5)) (\ep_5 \cdot k_3)(\ep_3 \cdot x_1) {\cal A}^{\rm het}( 1, 2, 4, 5, 3)   %ex[3, 1] ex[4, 1] ex[5, 1]
  \notag \\
%ek[5, 3] ex[3, 1] ex[4, 1] 
&\! \! \! \! \! \! \! \!  + (\ep_5\cdot k_3) (\ep_3 \cdot x_1)(\ep_4 \cdot x_1) {\cal A}^{\rm het}( 1, 4, 2,5, 3) + (\ep_3 \cdot x_1)(\ep_4 \cdot x_1) (\ep_5 \cdot x_1)
 {\cal A}^{\rm het}(1, 3, 4, 5, 2) 
% (ek[4, 3] + ek[4, 5]) ek[5, 3] ex[3, 1]
 \notag \\
%ee[3, 5] ek[4, 5] s[1, 3] 
 &\! \! \! \! \! \! \! \!  +\tfrac{1}{2} {\cal E}_{45}  (\ep_3 \cdot x_1)  
 \big[ s_{3 4} {\cal A}^{\rm het}( 3, 1, 2,5, 4)  - s_{1 4} {\cal A}^{\rm het} (1, 3, 2, 5, 4)  \big] \notag \\
 &\! \! \! \! \!  \! \! \! + \frac{2\alpha'  }{3 } \, {\cal F}_{345} \, s_{1 5} {\cal A}^{\rm het} (1, 2,3, 4, 5) +  {\rm perm}(3,4,5) \ .
  \end{align}
The field-theory limit of (\ref{threegr}) is obtained by again setting ${\cal E}_{pq} \rightarrow (\ep_p\cdot \ep_q),  \ {\cal A}^{\text{het}} (\ldots) \rightarrow A^{\text{YM}} (\ldots)$ and by discarding the subleading term $2 \alpha'{\cal F}_{pqr}$, in agreement with section 5.6 of \cite{Nandan:2016pya}.

%%%%%%%%%%%%%%%%%%%%%%
\section{Towards multitrace amplitudes} 
\label{section5}
%%%%%%%%%%%%%%%%%%%%%%

In this section, we reduce multitrace amplitudes (\ref{2.22}) involving $n$ gluons and at most one graviton to their single-trace counterparts in the gauge sector. The main challenge is posed by the conversion (\ref{2.17b}) of products of Parke-Taylor factors and graviton contributions to a single Parke-Taylor factor. This rearrangement will be addressed via integration by parts which resembles the application of scattering equations in the analogous CHY analysis \cite{Nandan:2016pya} but, as we shall see, organizes the low-energy regime in a different manner.

%%%%%%%%%%%%%%%%%%%%%%
\subsection{No graviton} 
\label{section51}
%%%%%%%%%%%%%%%%%%%%%%

In absence of gravitons, the general formula (\ref{2.22}) for double-trace amplitudes reduces to
\begin{align}
{\cal A}^{\rm het}_{(2)}( \{1,\ldots,r\, |\, r{+}1,\ldots,n\}) &= - \int_{\mathbb C^{n}} \frac{ \dd^2 z_1 \,  \ldots \, \dd^2 z_{n} }{ {\rm vol}({\rm SL}(2,\mathbb C))} \; {\cal K}_{n}(\{\bar z_j\}) \, {\cal C}_z(1,\ldots,r) \,  {\cal C}_z(r{+}1,\ldots,n)   \ .
\label{4.1}
\end{align}
As we will demonstrate below, suitable rearrangements of the Parke-Taylor factors on the right hand side via integration by parts lead to the following reduction to single-trace amplitudes:
\begin{align}
&{\cal A}^{\rm het}_{(2)}(\{ 1,2,\ldots,r \, | \,r{+}1,\ldots,n \}) =   -  \, \frac{2\alpha'}{1-2\alpha' s_{12\ldots r} }  \label{4.2}
 \\
& \ \ \ \ \  \times \sum_{j=1}^{r-1} \sum_{\ell=r+2}^{n} (-1)^{n-\ell} s_{j\ell}  \sum_{\tau \in \{ r+2,\ldots, \ell-1 \} \atop { \shuffle \{n,n-1,\ldots,\ell+1 \} } } \sum_{\sigma \in \{1,2,\ldots, j-1\} \atop{ \shuffle \{r+1,\tau,\ell \} }} \! \! {\cal A}^{\rm het}(\sigma,j,j{+}1,\ldots,r) \ .
\notag
\end{align}
Note that the sets $\tau$ from the first sum over shuffles (\ref{3.7}) enter the range of $\sigma$ in the second shuffle sum. With two, three and four particles in one of the traces, we have
\begin{align}
{\cal A}^{\rm het}_{(2)}(\{ 1,2,\ldots,r \, |\, p,q \}) &=
- \, \frac{2\alpha'}{1-2\alpha' s_{pq} }  \sum_{j=1}^{r-1} s_{jq} \! \! \!  \sum_{ \sigma \in \{ p,q\} \atop { \shuffle \{ 1,2,\ldots,j-1 \} } } \! \! \!  {\cal A}^{\rm het}(\sigma,j,j{+}1,\ldots,r) \label{4.3} \\
{\cal A}^{\rm het}_{(2)}(\{1,2,\ldots,r \, |\, p,q,t \}) &= \frac{2\alpha'}{1-2\alpha' s_{pqt} } 
 \sum_{j=1}^{r-1} s_{jq} \! \! \!  \sum_{ \sigma \in \{ p,t,q\} \atop { \shuffle \{ 1,2,\ldots,j-1 \} } } \! \! \!  {\cal A}^{\rm het}(\sigma,j,j{+}1,\ldots,r) - (q\leftrightarrow t)  \label{4.4} \\
{\cal A}^{\rm het}_{(2)}(\{ 1,2,\ldots,r  \, |\, p,q,t,u \}) &=  \frac{2\alpha'}{1-2\alpha' s_{pqtu} }  \sum_{j=1}^{r-1} \Big\{ s_{jt} \! \! \!  \sum_{ \sigma \in \{ p,q,u,t\} \atop { \shuffle \{ 1,2,\ldots,j-1 \} } }  \! \! \! {\cal A}^{\rm het}(\sigma,j,j{+}1,\ldots,r) \label{4.5} \\
&\ \ \ \ \ \  \ \ \ \ \ \ \ \ \ \ \ \ \ \ \  - s_{jq} \! \! \!  \sum_{ \sigma \in \{ p,u,t,q \} \atop { \shuffle \{ 1,2,\ldots,j-1 \} } } \! \! \!  {\cal A}^{\rm het}(\sigma,j,j{+}1,\ldots,r) \Big\} + (q\leftrightarrow u)  \ . \notag
\end{align}
At low multiplicity, examples of (\ref{4.3}) include
\begin{align}
{\cal A}^{\rm het}_{(2)}(\{ 1,2\, |\, 3,4 \}) &= -\, \frac{2\alpha' s_{14} }{1-2\alpha' s_{12}}  \, {\cal A}^{\rm het}(1,2,3,4) \label{4.6} \\
{\cal A}^{\rm het}_{(2)}(\{ 1,2,3\, |\, 4,5 \}) &= \frac{ 2\alpha' }{1-2\alpha' s_{45}}  \, \big[ s_{25} {\cal A}^{\rm het}(2,1,3,4,5) - s_{15} {\cal A}^{\rm het}(1,2,3,4,5) \big]  \ .\label{4.7} 
\end{align}
The right hand sides of the above relations do not manifest cyclicity within the traces or symmetry under their exchange. These properties can be checked to hold by means of BCJ relations (\ref{2.bcj}) of the single-trace amplitudes.

In the field-theory limit, double- and single-trace amplitudes of the heterotic string reduce to those of EYM and YM, respectively, i.e.\ ${\cal A}_{(2)}^{\rm het}(\{ \ldots \}) \rightarrow 2\alpha' A^{\rm EYM}(\{ \ldots \})$ as well as ${\cal A}^{\rm het}(\ldots) \rightarrow A^{\rm YM}(\ldots)$. The additional factor of $2\alpha'$ in the conversion between ${\cal A}_{(2)}^{\rm het} $ and $2\alpha' A^{\rm EYM}$ stems from the gravitational constant in the coupling between graviton and gluon which we suppress in the normalization conventions for $A^{\rm EYM}$ chosen in this work and \cite{Nandan:2016pya}. Then, the EYM relations of \cite{Nandan:2016pya} are recovered in the low-energy limit of (\ref{4.2}):
\begin{align}
&A^{\rm EYM}(\{ 1,\ldots,r \, | \,r{+}1,\ldots,n \}) =   -  \sum_{j=1}^{r-1} \sum_{\ell=r+2}^{n} (-1)^{n-\ell} s_{j\ell} \! \! \! \! \! \!  \sum_{\tau \in \{ r+2,\ldots, \ell-1 \} \atop { \shuffle \{n,n-1,\ldots,\ell+1 \} } } \sum_{\sigma \in \{1,2,\ldots, j-1\} \atop{ \shuffle \{r+1,\tau,\ell \} }} \! \!  \! \! \! A^{\rm YM}(\sigma,j,j{+}1,\ldots,r) \ .
\label{4.8}
\end{align}
The following subsections are devoted to the derivation of the all-multiplicity identity (\ref{4.2}).

%%%%%%%%%%%%%%%%%%%%%%
\subsubsection{Total derivatives} 
\label{section512}
%%%%%%%%%%%%%%%%%%%%%%

In this section, we generalize the total derivative (\ref{3.10}) which was used in (\ref{3.11}) to remove the Parke-Taylor factor ${\cal C}_z(p,q)$ from the two-graviton integrand. This will be done in
an ${\rm SL}(2,\mathbb C)$-frame where $z_r \rightarrow \infty$ and Koba-Nielsen derivatives simplify to $\frac{ \partial }{\partial z_i} {\cal I}_n = 2\alpha'{\cal I}_n \sum_{j\neq i,r} \frac{ s_{ij} }{z_{ij}} $ excluding $j=r$. We recapitulate the two-particle case (\ref{3.10}) for legs $1,2,\ldots,r$ and $p,q$ as
\beq
 \Big( \frac{1}{2\alpha' } \, \frac{\partial}{\partial z_q} + \sum_{j=1}^{r-1} \frac{ s_{jq} }{z_{j,q}} \Big) \, \frac{ {\cal I}_{r+2} }{z_{p,q}} =    (s_{pq}-\tfrac{1}{2\alpha'}) \,  {\cal C}_z(p,q) \, {\cal I}_{r+2}  \ ,
\label{4.11}
\eeq
and find analogous identities relevant to (\ref{4.4}) and (\ref{4.5}) when ${\cal C}_z(1,2,\ldots,r)$ is augmented by a second 
Parke-Taylor factor with legs $p,q,t$ and $p,q,t,u$,
\begin{align}
  (s_{pqt}-\tfrac{1}{2\alpha'})\, {\cal C}_z(p,q,t) \, {\cal I}_{r+3} &=  \Big( \frac{1}{2\alpha' } \,\frac{\partial}{\partial z_t} + \sum_{j=1}^{r-1} \frac{ s_{jt} }{z_{j,t}} \Big) \, \frac{ {\cal I}_{r+3} }{z_{p,q}z_{q,t}}  - (q\leftrightarrow t)  \label{4.12}
  \\
    (s_{pqtu}-\tfrac{1}{2\alpha'})\, {\cal C}_z(p,q,t,u) \, {\cal I}_{r+4} &=  \Big(\frac{1}{2\alpha' } \, \frac{\partial}{\partial z_u} + \sum_{j=1}^{r-1} \frac{ s_{ju} }{z_{j,u}} \Big) \, \frac{ {\cal I}_{r+4} }{z_{p,q}z_{q,t}z_{t,u}} \label{4.13} \\
    &  - \Big(\frac{1}{2\alpha' } \, \frac{\partial}{\partial z_t} + \sum_{j=1}^{r-1} \frac{ s_{jt} }{z_{j,t}} \Big) \, \frac{ {\cal I}_{r+4} }{z_{p,q}z_{q,u}z_{u,t}}   + (q\leftrightarrow u) \ . \notag
\end{align}
This generalizes as follows to arbitrary numbers of legs in both Parke-Taylor factors,
\begin{align}
 & (s_{12\ldots r}-\tfrac{1}{2\alpha'})\, {\cal C}_z(r{+}1,\ldots,n) \, {\cal I}_n =   \sum_{\ell=r+2}^n   (-1)^{n-\ell}  \label{4.14}  \\
 & \ \ \ \ \times  \Big( \frac{1}{2\alpha' } \, \frac{\partial}{\partial z_\ell} + \sum_{j=1}^{r-1} \frac{ s_{j\ell} }{z_{j,\ell}} \Big)   \! \! \! \! \sum_{\tau \in \{n,n-1,\ldots,\ell+1\}    \atop{ \shuffle \{r+2,r+3,\ldots,\ell-1\}}}   \frac{{\cal I}_n}{ z_{r+1,\tau_1}z_{\tau_1,\tau_2} \ldots z_{\tau_{|\tau|-1},\tau_{|\tau|}} z_{\tau_{|\tau|},\ell} }  \ ,
\notag
\end{align}
where $|\tau|$ denotes the cardinality of the set $\tau= \{\tau_1,\tau_2,\ldots,\tau_{|\tau|}\}$ and $s_{12\ldots r}=s_{r+1\ldots n}$.

%%%%%%%%%%%%%%%%%%%%%%
\subsubsection{Parke-Taylor rearrangements} 
\label{section513}
%%%%%%%%%%%%%%%%%%%%%%

Once we drop the total derivatives in the above identities and cast the contributions from $\frac{s_{j\ell} }{z_{j,\ell}}$ into
an ${\rm SL}(2,\mathbb C)$-frame independent form, (\ref{4.14}) allows to equate
\begin{align}
&\, (s_{12\ldots r} - \tfrac{1}{2\alpha'}) \, {\cal C}_z(1,2,\ldots,r ) \, {\cal C}_z(r{+}1,\ldots,n) =   - \sum_{j=1}^{r-1} \sum_{\ell=r+2}^{n} (-1)^{n-\ell} s_{j\ell} \! \! \
\notag \\
& \ \ \ \ \ \ \ \ \ \ \ \ \ \ \ \ \ \ \ \ \ \ \ \ \times \sum_{\tau \in \{ r+2,\ldots, \ell-1 \} \atop { \shuffle \{n,n-1,\ldots,\ell+1 \} } } \sum_{\sigma \in \{1,2,\ldots, j-1\} \atop{ \shuffle \{r+1,\tau,\ell \} }} \! \! {\cal C}_z(\sigma,j,j{+}1,\ldots,r) 
\label{4.15}
\end{align}
upon integration against the Koba-Nielsen factor. The restrictions of (\ref{4.15}) to two, three and four legs in the second Parke-Taylor factor are obtained from (\ref{4.11}) to (\ref{4.13}) and read
\begin{align}
(s_{pq} - \tfrac{1}{2\alpha'})\, {\cal C}_z(1,2,\ldots,r ) \, {\cal C}_z(p,q) &=
- \sum_{j=1}^{r-1} s_{jq} \! \! \!  \sum_{ \sigma \in \{ p,q\} \atop { \shuffle \{ 1,2,\ldots,j-1 \} } } \! \! \!  {\cal C}_z(\sigma,j,j{+}1,\ldots,r) \label{4.17} \\
(s_{pqt} - \tfrac{1}{2\alpha'})\, {\cal C}_z(1,2,\ldots,r)\, {\cal C}_z(p,q,t) &=
 \sum_{j=1}^{r-1} s_{jq} \! \! \!  \sum_{ \sigma \in \{ p,t,q\} \atop { \shuffle \{ 1,2,\ldots,j-1 \} } } \! \! \!  {\cal C}_z(\sigma,j,j{+}1,\ldots,r) - (q\leftrightarrow t) \label{4.18}  \\
(s_{pqtu} - \tfrac{1}{2\alpha'})\,  {\cal C}_z(1,2,\ldots,r ) \, {\cal C}_z(p,q,t,u) &= \sum_{j=1}^{r-1} \Big\{ s_{jt} \! \! \!  \sum_{ \sigma \in \{ p,q,u,t\} \atop { \shuffle \{ 1,2,\ldots,j-1 \} } }  \! \! \! {\cal C}_z(\sigma,j,j{+}1,\ldots,r) \label{4.19}  \\
&\ \ \ \ \ \  - s_{jq} \! \! \!  \sum_{ \sigma \in \{ p,u,t,q \} \atop { \shuffle \{ 1,2,\ldots,j-1 \} } } \! \! \!  {\cal C}_z(\sigma,j,j{+}1,\ldots,r) \Big\} + (q\leftrightarrow u)  \ , \notag
\end{align}
see (\ref{3.75}) for an earlier application of (\ref{4.18}). These integrations by parts can be viewed as the string-theory counterparts of the cross-ratio relations of \cite{Cardona:2016gon}, also see \cite{Cachazo:2015nwa, Gomez:2016bmv} for related work on CHY integrals involving multiple Parke-Taylor factors. Note that (\ref{4.15}) is instrumental for the reduction of disk integrals to an $(n{-}3)!$ basis in the open superstring \cite{Mafra:2011nv, Mafra:2011nw} and the open bosonic string \cite{Huang:2016tag}. However, as exemplified by the following section, (\ref{4.15}) is not yet sufficient to address the most general integrand with suitable ${\rm SL}(2,\mathbb C)$ weights.

%%%%%%%%%%%%%%%%%%%%%%
\subsubsection{Comparison with CHY and breakdown of uniform transcendentality} 
\label{section515}
%%%%%%%%%%%%%%%%%%%%%%

The above integration by parts identities (\ref{4.15}) which reduce products of two Parke-Taylor factors to a single one are seen to introduce factors of $(1-2\alpha' s_{12\ldots r})^{-1}$ into the amplitude relation (\ref{4.2}). It is instructive to compare this with the EYM integrands for gluonic double-trace amplitudes in the CHY formalism \cite{Cachazo:2014nsa, Cachazo:2014xea}. These CHY integrands augment the two Parke-Taylor factors by an explicit numerator factor of $s_{12\ldots r}$ which can be thought of as compensating the $\alpha'$-dependent part of the string-theory specific denominator $(1-2\alpha' s_{12\ldots r})^{-1}$: A sequence of scattering equations \cite{Cardona:2016gon, Nandan:2016pya} only yields $s_{12\ldots r}^{-1}$ in the place of $( s_{12\ldots r} - \frac{1}{2\alpha'})^{-1}$ when converting a product of two Parke-Taylor factors to a single one. In the field-theory limit, the net effect of $(1-2\alpha' s_{12\ldots r})^{-1} \rightarrow 1$ is the same as the conspiration between the outcome $\sim s_{12\ldots r}^{-1}$ of scattering equations and the additional numerator $s_{12\ldots r}$ in the CHY integrand. That is why the heterotic string and the CHY formalism converge on the EYM amplitude relations (\ref{4.8}), although the opening line for their integrands differ by $s_{12\ldots r} $.

Similar to the discussion in section \ref{section413}, the tentative tachyon poles of $(1-2\alpha' s_{12\ldots r})^{-1}$ at $(k_{1}{+}\ldots{+}k_{r})^2=\alpha'$ are cancelled by the zeros of the accompanying worldsheet integrals in the single-trace amplitudes, in lines with level matching. A geometric-series expansion of the denominator factors shows explicitly that double-trace amplitudes of gluons do not exhibit uniform transcendentality. Since the complexity of the integrations by parts and the associated denominators grows with the number of traces and graviton insertions (see the next section) we expect that uniform transcendentality is violated in {\em any} multitrace amplitude of the heterotic string.

%%%%%%%%%%%%%%%%%%%%%%
\subsection{One graviton} 
\label{section52}
%%%%%%%%%%%%%%%%%%%%%%

In presence of a single graviton, the general formula (\ref{2.22}) for double-trace amplitudes reduces to
\begin{align}
{\cal A}^{\rm het}_{(2)}( \{1,2,\ldots,r\, |\, r{+}1,\ldots,n\};p) &= - \int_{\mathbb C^{n+1}} \! \! \! \! \frac{ \dd^2 z_1 \,  \ldots \, \dd^2 z_{n} \, \dd^2 z_p  }{ {\rm vol}({\rm SL}(2,\mathbb C))} \; {\cal K}_{n+1}(\{\bar z_j\}) \notag \\
&\! \! \! \! \! \! \! \! \! \! \! \! \! \! \! \! \! \! \! \! \! \! \! \! \!  \! \! \! \! \! \! \! \! \! \!  \times \, {\cal C}_z(1,2,\ldots,r) \,  {\cal C}_z(r{+}1,\ldots,n)  \, R_n(\ep_p;\{k_j,z_j\}) \ .
\label{4.21}
\end{align}
It will be elaborated in the following paragraphs that suitable integrations by parts combine the Parke-Taylor factors in the second line with the graviton contribution $R_n(\ep_p;\{k_j,z_j\}) $ given in (\ref{2.23}) such that
\begin{align}
&\mathcal{A}^{\rm het}_{(2)}(\{1,2,\ldots,r \, | \,r{+}1, \ldots,n\}; p)= \sum_{l=1}^{r-1} (\ep_p\cdot x_l)  
	 \, \mathcal{A}^{\rm het}_{(2)}(\{ 1,2,\ldots,l,p,l{+}1,\ldots,r\, | \,r{+}1, \ldots,n \}) \notag \\
 & \ \ \ \ \ \ \ \ \    +\sum_{l=r+1}^{n-1}   (\ep_p\cdot x_l)  
	\, \mathcal{A}^{\rm het}_{(2)}(\{ 1,2,\ldots,r\, | \,r{+}1, \ldots,l,p,l{+}1,\ldots,n \}) \ ,  \label{4.41} \\
& \ \ \ \ \ \ \ \ \  - \, \frac{2\alpha'\, (\ep_p\cdot x_r)}{1-2\alpha' s_{12\ldots r}} \, \Big\{  \sum_{i=1}^{r-1} \sum_{j=r+2}^{n} (-)^{i-j} s_{ij} \! \! \! \! \! \sum_{\sigma  \in \{1,2,\ldots, i-1\} \atop{ \shuffle \{r-1,r-2,\ldots, i+1\} }} \sum_{\tau \in \{  j-1,j-2,\ldots, r+2\} \atop{ \shuffle \{j+1,j+2,\ldots, n \}  }} \! \! \! \! \!
 {\cal A}^{\rm het}(r, \sigma,i,j,\tau, r{+}1 ,p)    \notag \\
 & \ \ \ \ \ \ \ \ \ \  \ \ \ \ \ \ \ \ \ \  \ \ \ \ \ \ \ \ \ \ \ \ \ \ \ \  +  \sum_{l=1}^{r-1} (p\cdot x_l)  
	 \, \mathcal{A}^{\rm het}_{(2)}(\{ 1,2,\ldots,l,p,l{+}1,\ldots,r\, | \,r{+}1, \ldots,n \}) \Big\}  \ .\notag
\end{align}
The right hand side is presented in terms of gluon amplitudes of both single- and double-trace type, where the latter may also be reduced to the single-trace sector via (\ref{4.2}). In the simplest setting with two gluons in both traces, (\ref{4.41}) specializes to
\begin{align}
\mathcal{A}^{\rm het}_{(2)}(\{1,2 \, | \,3,4\}; p) &= (\ep_p \cdot x_1) \, \mathcal{A}^{\rm het}_{(2)}(\{1, p, 2\,|\, 3, 4\}) + 
(\ep_p \cdot x_3) \,\mathcal{A}^{\rm het}_{(2)}(\{1, 2 \,|\,  3, p, 4\}) \notag \\
& \! \! \!  \! \! \!  \! \! \!  \! \! \!  \! \! \!  \! \! \!  \! \! \!  \! \! \!  \! \! \!   + 
 \frac{2\alpha' \, (\ep_p \cdot x_2) }{1-2\alpha' \,s_{12}} \, \Big\{  s_{14}  \mathcal{A}^{\rm het}({2, 1, 4, 3, p}) - s_{15} \mathcal{A}^{\rm het}_{(2)}(\{1, p, 2\,|\, 3, 4\}) \Big\} \ .  \label{4.42}
\end{align}
In taking the field-theory limit of (\ref{4.41}), we note that its first two lines $\sim  \mathcal{A}^{\rm het}_{(2)} $ are of the same order in $\alpha'$ as the single-trace expressions $\sim 2\alpha' s_{ij} {\cal A}^{\rm het}$ in the third line. The geometric-series expansion of $(1-2\alpha' s_{12\ldots r})^{-1}$ and the entire fourth line, however, are subleading in $\alpha'$ and decouple from the low-energy limit. We therefore reproduce the field-theory relation of \cite{Nandan:2016pya} from the leading $\alpha'$-order of (\ref{4.41}) with $ \mathcal{A}^{\rm het}_{(2)} \rightarrow 2\alpha' A^{\rm EYM}$:
\begin{align}
& A^{\rm EYM}(\{1,2,\ldots,r \, | \,r{+}1, \ldots,n\}; p)= \sum_{l=1}^{r-1} (\ep_p\cdot x_l)  
	 \, A^{\rm EYM}( \{ 1,\ldots,l,p,l{+}1,\ldots,r\, | \,r{+}1, \ldots,n \}) \notag \\
 & \ \ \ \ \ \ \ \ \ \ \   +\sum_{l=r+1}^{n-1}   (\ep_p\cdot x_l)  
	\, A^{\rm EYM}(\{1,\ldots,r\, | \,r{+}1, \ldots,l,p,l{+}1,\ldots,n \})   \label{4.2ft}  \\
& \ \ \ \ \ \ \ \ \ \ \   - \, (\ep_p\cdot x_r) \, \sum_{i=1}^{r-1} \sum_{j=r+2}^{n} (-)^{i-j} s_{ij} \sum_{\sigma  \in \{1,2,\ldots, i-1\} \atop{ \shuffle \{r-1,r-2,\ldots, i+1\} }} \sum_{\tau \in \{  j-1,j-2,\ldots, r+2\} \atop{ \shuffle \{j+1,j+2,\ldots, n \}  }}
 A^{\rm YM}(r, \sigma,i,j,\tau, r{+}1 ,p)    \ . \notag 
\end{align}
We shall now turn to the derivation of (\ref{4.41}).

%%%%%%%%%%%%%%%%%%%%%%
\subsubsection{The structure of the graviton correlator} 
\label{section521}
%%%%%%%%%%%%%%%%%%%%%%

The representation (\ref{3.1}) of the graviton correlator $R_n(\ep_p;\{k_j,z_j\}) $ yields three classes of terms
\beq
R_n(\ep_p;\{k_j,z_j\})= (\ep_p \cdot x_r) \,   \frac{ z_{r,r+1} }{z_{r,p} z_{p,r+1}}  +
\sum_{j=1}^{r-1} (\ep_p \cdot x_j) \, \frac{ z_{j,j+1} }{z_{j,p} z_{p,j+1}}
+\sum_{j=r+1}^{n-1} (\ep_p \cdot x_j) \,\frac{ z_{j,j+1} }{z_{j,p} z_{p,j+1}}   \ ,
\label{4.22}
\eeq
where the $j$-dependent combinations $\frac{ z_{j,j+1} }{z_{j,p} z_{p,j+1}}$ signal the insertion of the graviton leg $p$ into the Parke-Taylor factors of (\ref{4.21}), i.e.
\begin{align}
&{\cal C}_z(1,2,\ldots,r) \,  {\cal C}_z(r{+}1,\ldots,n)\,  R_n(\ep_p;\{k_j,z_j\}) =  (\ep_p \cdot x_r) \, {\cal C}_z(1,2,\ldots,r) \,  \frac{ z_{r,r+1} }{z_{r,p} z_{p,r+1}}   \, {\cal C}_z(r{+}1,\ldots,n) 
\notag \\
& \ \ \ \ \ \ \ \  \ \ \ \ \ \ \ \  \ \ \ \ \ \ \ \  \ \ \ \ \ \ \ \  + {\cal C}_z(1,2,\ldots,r) \sum_{j=r+1}^{n-1} (\ep_p \cdot x_j) \, {\cal C}_z(r{+}1,\ldots,j,p,j{+}1,\ldots,n)
\label{4.23} \\
& \ \ \ \ \ \ \ \  \ \ \ \ \ \ \ \  \ \ \ \ \ \ \ \  \ \ \ \ \ \ \ \  +   {\cal C}_z(r{+}1,\ldots,n) \sum_{j=1}^{r-1} (\ep_p \cdot x_j) \, {\cal C}_z(1,2,\ldots,j,p,j{+}1,\ldots,r) \ .
\notag
\end{align}
While the sums over $j$ in (\ref{4.23}) directly translate into the first and second line of (\ref{4.41}), the term $\sim (\ep_p \cdot x_r)$ is not yet in a suitable form to make contact with any of ${\cal A}^{\rm het}(\ldots)$ or ${\cal A}^{\rm het}_{(2)}( \{ \ldots\})$.

%%%%%%%%%%%%%%%%%%%%%%
\subsubsection{A new class of integrations by parts} 
\label{section522}
%%%%%%%%%%%%%%%%%%%%%%

The leftover task is to express the first line of (\ref{4.23}) $\sim (\ep_p \cdot x_r)$ in Parke-Taylor form, where contributions with one or two factors of ${\cal C}_z(\ldots)$ are equally acceptable in view of (\ref{4.2}). It turns out that a sequence of integrations by parts along the lines of section \ref{section512} yields
\begin{align}
&(s_{12\ldots r}-\tfrac{1}{2\alpha'})  \, {\cal C}_z(1,2,\ldots,r) \, \frac{ z_{r,r+1} }{z_{r,p} z_{p,r+1}}   \, {\cal C}_z(r{+}1,\ldots,n)
 \notag \\
 & \ \ \ \  = \sum_{l=1}^{r-1} (p\cdot x_l) \, {\cal C}_z(1,2,\ldots,l,p,l{+}1,\ldots,r)  \, {\cal C}_z(r{+}1,\ldots,n) \label{4.24}  \\
 & \ \ \ \ \ \ \ \ \ 
 - \sum_{i=1}^{r-1} \sum_{j=r+2}^n (-1)^{i-j} s_{ij} \! \! \! \! \! \sum_{\sigma  \in \{1,2,\ldots, i-1\} \atop{ \shuffle \{r-1,r-2,\ldots, i+1\} }} \sum_{\tau \in \{  j-1,j-2,\ldots, r+2\} \atop{ \shuffle \{j+1,j+2,\ldots, n \}  }} \! \! \! \! \! {\cal C}_z(r,\sigma,i,j,\tau,r+1,p) \notag
\end{align}
when integrated against the $(n{+}1)$-particle Koba-Nielsen factor in (\ref{4.21}).
The antisymmetry of the left hand side under exchange of $(r,1,2,\ldots,r{-}1)\leftrightarrow (r{+}1,\ldots,n)$ is not reflected in the presentation of the right hand side. Accordingly, the contributions of (\ref{4.24}) to the amplitude relation (\ref{4.41}) --
specifically its third and fourth line -- do not manifest the symmetry under exchange of the gluon traces. Still, the hidden exchange symmetry can be verified by expanding the double-trace amplitudes in a single-trace basis via (\ref{4.2}) and applying BCJ relations (\ref{2.bcj}).

%%%%%%%%%%%%%%%%%%%%%%
\subsubsection{Comparison with CHY} 
\label{section526}
%%%%%%%%%%%%%%%%%%%%%%

The CHY integrand for multitrace EYM amplitudes with any number of graviton insertions has been given in \cite{Cachazo:2014xea}. The string-theory correlator $R_n(\ep_p;\{k_j,z_j\})$ for a single graviton was observed earlier on to match with the diagonal elements $C_{pp}$ of the $C$-matrix in the reduced Pfaffian \cite{Cachazo:2013hca}. Indeed, $C_{pp}$ appears in the double-trace CHY integrand for one graviton\cite{Cachazo:2014xea}, but it is accompanied by a factor of $s_{12\ldots p}$ as well as extra summands spelt out in \cite{Nandan:2016pya}. It turns out that the additional ingredients of the CHY integrand besides $R_n(\ep_p;\{k_j,z_j\})$ compensate for the different organizations of the integral over ${\cal C}_z(1,2,\ldots,r) \, \frac{ z_{r,r+1} }{z_{r,p} z_{p,r+1}}   \, {\cal C}_z(r{+}1,\ldots,n)$ in the two setups.

In string theory, the products of Parke-Taylor factors in the middle line of the integration by parts identity (\ref{4.24}) are subleading in $\alpha'$ and do not contribute to the EYM relation (\ref{4.2ft}). The counterpart of this middle line in the analogous scattering equations, however, does contribute to the field-theory order and cancels some of the extra terms of the CHY integrand which do not appear in the string-theory correlator. This example gives rise to suspect that a systematic handle on string integrals with multiple Parke-Taylor factors and additional cross ratios as in (\ref{4.24}) might allow for a streamlined approach to EYM relations involving multiple traces, without the need to track spurious sectors of the CHY integrands.

%%%%%%%%%%%%%%%%%%%%%%
\section{Conclusions} 
\label{section6}
%%%%%%%%%%%%%%%%%%%%%%

We have presented novel relations between tree-level amplitudes in heterotic string theory involving gluons and gravitons, including their supersymmetry multiplets and $\alpha'$-corrections. Purely gluonic single-trace amplitudes were identified as the fundamental building blocks to capture both single- and double-trace amplitudes with additional graviton insertions and an arbitrary number of gluons in each trace. The derivation of such amplitude relations is based on algebraic rearrangements of the worldsheet integrands and integration by parts. These techniques have led us to explicit results for single-trace amplitudes with up three gravitons in (\ref{1.1}), (\ref{twograv}) and (\ref{threegr}) as well as double-trace amplitudes with at most one graviton, see (\ref{4.2}) and (\ref{4.41}). 

The algorithmic derivation of the results in this work gives rise to expect that any multitrace amplitude with an arbitrary number of graviton insertions can be reduced to single-trace amplitudes among gluons by iterating the above techniques\footnote{This amounts to rewriting any correlation functions with Kac-Moody currents and graviton insertions in terms of Parke-Taylor factors. A recursive solution to the analogous problem in the CHY formalism is given in \cite{Cardona:2016gon}, where integrations by parts are replaced by scattering equations, at the expense of shifting some of the expansion coefficients $s_{ij\ldots k}$ by $\frac{1}{2\alpha'}$.}. Instead of performing the tedious but by now straightforward analysis for additional multigraviton and multitrace cases and displaying the resulting formulae of increasing length, we hope to address the general problem through a recursion in the number of gravitons and traces in the near future.

Our results extend the connections between tree-level amplitudes in different string vacua found in \cite{Stieberger:2014hba}. In this reference, gluonic single-trace amplitudes in heterotic string theory were related to single-valued \cite{Schnetz:2013hqa, Brown:2013gia} gluon trees of the type-I as well as graviton trees of the type-II superstring. On these grounds, our findings pave the way to reducing the entire tree-level S-matrix for gluons {\em and} gravitons in heterotic string theory to purely gluonic type-I input, augmented by the additional polarization vectors from the gravitons. This goes beyond the heterotic-type-I duality \cite{Polchinski:1995df} and might carry some general lessons on the web of string dualities where each theory is interpreted as a separate vacuum of an overarching M-theory \cite{Witten:1995ex}.

Finally, it would be interesting to directly compare the $\alpha'$-expansions of single-trace amplitudes involving the {\em same} gluon and graviton states between the heterotic string and the type-I superstring. However, tentative matchings between the heterotic-string results of this work and type-I expressions in \cite{Stieberger:2009hq, Stieberger:2015vya, Stieberger:2016lng} can only occur in specific subsectors of their $\alpha'$-expansion: the single-valued image of the type-I amplitudes and the uniform-transcendentality parts\footnote{Note that uniform transcendentality has been recognized as a universality criterion for open-string interactions and purely gravitational closed-string interactions in different string theories \cite{Huang:2016tag}.} on the heterotic side. Uniform transcendentality refers to the correspondence between powers of $\alpha'$ and the weights of the accompanying MZVs, and we observe it to be violated by multitrace amplitudes and by single-trace amplitudes with two or more graviton states in heterotic string theory. 

In summary, this work contributes new structural insights into the tree-level S-matrix of string theories and 
is expected to open the door to numerous further research directions.

%%%%%%%%%%%%%%%%%%%%%%
\subsubsection*{Acknowledgments}
%%%%%%%%%%%%%%%%%%%%%%

I am grateful to Valentin Verschinin for enlightening discussions, Carlos Mafra, Dhritiman Nandan, Stefan Theisen and Congkao Wen for valuable comments on the draft as well as Yu-tin Huang, Dhritiman Nandan, Jan Plefka and Congkao Wen for fruitful collaboration on related topics. Moreover, I would like to thank the Yau Mathematical Sciences Center, Tsinghua University for kind hospitality during final stages of this project.

%\bibliographystyle{nb}
%\bibliography{gluongravitonbib}

%bibliography generated by nb.bst v1.06 (C) 2003-2011 Niklas Beisert

\end{document}